\documentclass[amsmath,amssymb,aps,prb,10pt,superscriptaddress,english,twocolumn,longbibliography]{revtex4-2}

\usepackage{amsmath,amssymb}
\usepackage{graphicx,epsfig}
\usepackage{dcolumn}
\usepackage{color,epstopdf}
\usepackage{tikz}

\usepackage{float}
\usepackage{bm}
\usepackage[normalem]{ulem}
\usepackage[colorlinks=true,bookmarks=false,citecolor=blue,linkcolor=blue]{hyperref}

\usepackage{multirow}

\usepackage{braket}
\usepackage{refmathstyle}
\usepackage{xcolor}

\def\ff{f}
\def\phat{\hat{p}}
\def\fg{g}
\def\fgg{G}
\def\fh{h}
\def\fhh{H}
\def\fea{e_1}
\def\feb{e_2}
\def\caln{\mathcal{N}}

\DeclareMathOperator\erf{Erf}
\DeclareMathOperator\tr{tr}

\DeclareMathOperator{\erfc}{Erfc}

\definecolor{dkgreen}{rgb}{0,0.6,0}

\definecolor{blue-green}{rgb}{0.0, 0.87, 0.87}

\begin{document}

\title{Probing symmetries of quantum many-body systems through gap ratio statistics}

\author{Olivier Giraud}
\email{olivier.giraud@lptms.u-psud.fr}
\affiliation{Universit\'e Paris-Saclay, CNRS, LPTMS, 91405, Orsay, France}
\author{Nicolas Mac\'e}
\email{mace@irsamc.ups-tlse.fr}
\affiliation{Laboratoire de Physique Th\'eorique, IRSAMC, Universit\'e de Toulouse, CNRS, UPS, France}
\author{\'Eric Vernier}
\email{vernier@lpsm.paris}
\affiliation{CNRS \& LPSM, Universit\'e Paris Diderot, place Aur\'elie Nemours, 75013 Paris, France}

\author{Fabien Alet}
\email{alet@irsamc.ups-tlse.fr}
\affiliation{Laboratoire de Physique Th\'eorique, IRSAMC, Universit\'e de Toulouse, CNRS, UPS, France}

\date{\today}

\begin{abstract}
The statistics of gap ratios between consecutive energy levels is a widely used tool, in particular in the context of many-body physics, to distinguish between chaotic and integrable systems, described respectively by Gaussian ensembles of random matrices and Poisson statistics. 
In this work we extend the study of the gap ratio distribution $P(r)$ to the case where discrete symmetries are present. This is important, since in certain situations it may be very impractical, or impossible, to split the model into symmetry sectors, let alone in cases where the symmetry is not known in the first place. 
Starting from the known expressions for surmises in the Gaussian ensembles, we derive analytical surmises for random matrices comprised of several independent blocks. We check our formulae against simulations from large random matrices, showing excellent agreement. We then present a large set of applications in many-body physics, ranging from quantum clock models and anyonic chains to periodically-driven spin systems. In all these models the existence of a (sometimes hidden) symmetry can be diagnosed through the study of the spectral gap ratios, and our approach furnishes an efficient way to characterize the number and size of independent symmetry subspaces. 
We finally discuss the relevance of our analysis for existing results in the literature, as well as its practical usefulness, and point out possible future applications and extensions.
\end{abstract}

\maketitle

\section{Introduction}
\label{sec:intro}

Symmetry considerations are an essential part of a physicist's toolbox, with countless applications in all fields of physics, ranging from Noether's theorem, gauge theories or the description of phase transitions~\cite{Gross_symmetry_1996}. 
Another frequent tool is the use of simplified models, which successfully describe the important features of a physical phenomenon without having to deal with all microscopic details. In this respect, Random Matrix Theory (RMT), which was first initiated to understand the statistical properties of energy levels in complex nuclei~\cite{wigner_1955}, is an extremely successful approach which has also impacted various branches in physics \cite{guhr1998random,mehta2004random}. It then comes as no surprise that symmetry properties are an integral part of RMT: one of the best-known examples is the construction of classical Gaussian ensembles from time-reversal symmetry considerations. Depending on the underlying symmetry of the system considered, it is best described by random matrices belonging to one of the three following ensembles: Gaussian Orthogonal Ensemble (GOE), Gaussian Unitary Ensemble (GUE) and Gaussian Symplectic Ensemble (GSE), whose entries are respectively real, complex or quaternionic random variables. Convenient to the description of Floquet
operators are the circular ensembles introduced by Dyson \cite{dyson1962statistical}: circular Orthogonal Ensemble (COE), circular Unitary Ensemble (CUE) and circular Symplectic Ensemble (CSE). These ensembles have the same asymptotic level spacing distributions as the Gaussian ensembles \cite{mehta2004random}.

The celebrated conjectures of Berry and Tabor~\cite{berry_1977} and Bohigas, Giannoni and Schmit~\cite{bohigas_1984} state that RMT describes the spectral statistics of quantum systems with a chaotic semiclassical limit, whereas Poisson statistics provides a description of systems with a classical integrable limit. These two paradigms serve as reference points to study the transitions between localization and ergodicity, for instance the Anderson transition as a function of disorder~\cite{evers_anderson_2008}. Quite crucially, quantum many-body systems, for which there is in general no semiclassical limit, also display the same dichotomy: RMT statistics for chaotic systems and Poisson statistics for quantum integrable systems, including those showing an emergent integrability such as many-body localized systems~\cite{pal_many_2010, alet_many_2018}. Numerous examples illustrate the usefulness of a RMT analysis of quantum many-body spectra~\cite{montambaux_1993,hsu_1993,bruus_1997,oganesyan_localization_2007,pal_many_2010}.

A universal tool in this respect is the study of the distribution $p(s)$ of level spacings, or gaps, defined as the differences between consecutive energy levels, $s_i=\lambda_i-\lambda_{i-1}$, assuming that the mean level density is fixed to unity, i.e.~$\langle s\rangle=1$. RMT offers simple, powerful predictions for the distribution $p(s)$ in terms of three different Wigner surmises corresponding to the three Gaussian ensembles mentioned above~\cite{wigner_1955}. These surmises are obtained by a simple calculation on random $2\times 2$ matrices, and turn out to reproduce most of the features of much larger random matrices, with high precision~\cite{dietz1990taylor}. Normalizing the level spacing distribution requires the knowledge of the density of states, which is often not analytically available. Numerically, one needs to perform an unfolding of the spectrum, for which there exists different procedures~\cite{haake_book,berry_1977,bruus_1997}. Unfolding can lead to spurious results~\cite{gomez_misleading_2002}, in particular because of finite-size effects; one may even find instances where different unfolding procedures leads to different physical interpretations of the same data. For many-body systems, the density of states is generically far from being uniform, which makes the use of the unfolding procedure rather inaccurate.

A very useful alternative to the study of $s$ has been proposed by Oganesyan and Huse~\cite{oganesyan_localization_2007}, in terms of the gap ratio for three consecutive levels, $r_i=\frac{\min(s_i,s_{i+1})}{\max(s_i,s_{i+1})}$. The key point is that considering the ratio of gaps rather than the gaps themselves suppresses the need to know or estimate the density of states, and thus avoids the numerical unfolding step. The probability distribution $P(r)$ of gap ratios is thus well-suited  to characterize statistical properties of many-body  spectra. The Poisson statistics distribution $P_{\rm Poisson}(r)=2/(1+r)^2$ can be easily derived from a Poisson sequence. For the random matrix spectra, analytical surmises of $P_{\rm GOE}(r), P_{\rm GUE}(r)$ and $P_{\rm GSE}(r)$ have been obtained by Atas {\it et al.}~\cite{atas2013distribution} from the joint eigenvalue distribution of $3\times 3$ random matrices, and improved estimates were obtained in \cite{atas2013joint} based on $4\times 4$ matrices.

Because of its computational advantage (no unfolding needed) and the existence of these analytical predictions, the gap ratio $r$, in particular its average $\langle r \rangle$ and its distribution $P(r)$, has become one of the most studied metrics in the field of disordered quantum systems. For instance, it is often used to characterize the change of statistics across a many-body localization (MBL) transition, between an ergodic phase, for which the RMT predictions for $P(r)$ are expected, and a Many-Body Localized phase, which displays emergent integrability and thus  $P_{\rm Poisson}(r)$ gap ratio statistics~\cite{pal_many_2010,cuevas_level_2012,luitz_many_2015}. The agreement between the RMT-predicted $P(r)$ and the numerical estimate for a given model now routinely diagnoses quantum chaotic models. Any discrepancy in the gap ratio as a function of a model parameter is often interpreted as a sign of a different physical behavior (see e.g.~\cite{khemani_signatures_2019}). The distribution of gap ratios is also instrumental in analyzing the symmetry properties of the SYK model and variants as a function of the number of Majorana fermions~\cite{you_SYK_2017, Kanazawa_2017,Li_supersymmetric_2017,iyoda_effective_2018,sun_periodic_2020}. $P(r)$ has also been measured experimentally to probe an ergodic to MBL transition / crossover~\cite{roushan_spectroscopic_2017}.  Applications of this metrics were also performed in other fields of study, such as in astrophysics~\cite{evano_correlations_2019}, for statistics of the zeros of the Riemann zeta function~\cite{atas2013distribution}, or characterizing entanglement in quantum circuits~\cite{Shaffer_2014}. 
The computation of the gap ratio statistics has been extended in several ways, such as ratios of gaps for levels with one or more other levels in-between, or non-Hermitian matrices~\cite{atas2013joint,Srivastava_2018,tekur_higher_2018,bhosale_scaling_2018,bhosale2019superposition,tekur2018symmetry,tekur_exact_2018,sa_complex_2020}.

What happens to spectral statistics in the situation where symmetries are present in the original Hamiltonian? In a seminal work~\cite{Rosenzweig_1960}, Rosenzweig and Porter computed the level spacing distribution $P(s)$ of systems with several independent random blocks (each being a random matrix with spacing distribution $p(s)$). This situation typically occurs when a physical system displays discrete symmetries, in which case the number of blocks remains finite in the thermodynamic limit. For continuous symmetries, the number of blocks grows with the system size, and ultimately, as many independent spectra are mixed, one expects a Poisson distribution to emerge for large enough systems. In an extension of the original work~\cite{Rosenzweig_1960}, Berry and Robnik considered mixed phase spaces with both ergodic and integrable blocks~\cite{Berry_1984}.

In general, one would be inclined to resolve the underlying symmetries by treating each block independently and performing a block diagonalization. This is not always possible. First, there are cases where a symmetry not previously known or analyzed is discovered fortuitously (e.g. by monitoring the gap ratio and seeing that it does not converge to its expected value). Second, in some situations, the block diagonalization is not known, too complex to implement, or cannot be performed totally. The latter case occurs for instance in systems with non-Abelian discrete symmetries, where two symmetry operations that commute with the Hamiltonian do not commute with each other (see e.g.~\cite{friedman_localization_2018,prakash_eigenstate_2017}). Third, there are cases where the basis transformation leading to a block structure in the Hamiltonian is known, but results in a Hamiltonian which is more costly to analyze; this is for instance the case if a sparse Hamiltonian leads to non-sparse blocks, inducing a strong decrease in the performances of numerical routines (we will present such an example in Sec.~\ref{sec:rsos}). 

In this work, we extend the Rosenzweig-Porter analysis to the computation of the gap ratio statistics $P(r)$ when several independent blocks are present. We do so by calculating the joint gap distribution $P(x,y)$ for a matrix with several independent random blocks, each being a random matrix with joint gap distribution $p(s,t)$. We obtain closed expressions for $P(x,y)$ and $P(r)$ in terms of $p(s,t)$ and its primitives. These expressions are valid for an arbitrary number of blocks. In the case of Gaussian random matrices, we use for $p(s,t)$ a surmise given by the exact $3\times 3$ distribution of RMT, which allows us to obtain expressions for $P(r)$. However our formula applies for an arbitrary distribution $p(s,t)$. We note that a recent work~\cite{sun_color_2020} provides estimates for $P(r)$ and $\langle r\rangle$ based on an surmise obtained from explicit analytical calculations for small-size matrices; however it does not take into account all possible level partitions. Our approach is quite different, as we discuss below. 

Our analytical estimates are virtually indistinguishable from numerical simulations on large random matrices. Our results explain several deviations for the distribution $P(r)$ or expectation value $\langle r \rangle$ observed in the literature, as discussed in Sec.~\ref{sec:literature}. They can also be useful in several situations such as those mentioned above (which we illustrate with various applications taken from many-body physics in Sec.~\ref{sec:applications}), as well as to estimate the number of effective ergodic blocks in an incompletely thermalized system. 

The manuscript is organized as follows. We first introduce the problem in Sec.~\ref{sec:problem}, setting up the notations and summarizing the useful literature as well as our own results. Sec.~\ref{sec:analytics1} contains the  derivation of the generic form of $P(r)$ when several independent blocks are present. We then present results for the three Gaussian ensembles. Sec.~\ref{sec:numerics} compares these analytically obtained results to simulations performed on random matrices, showing an excellent agreement. Sec.~\ref{sec:applications} contains several realistic applications of these results in many-body physics, with a panel of different types of possible symmetries: clock symmetries, symmetries in disorder realizations, dynamical symmetries in Floquet systems, disordered anyonic chains with topological symmetries. In Sec.~\ref{sec:discussion}, we finally conclude by first discussing existing examples where our work directly applies, and then suggesting some further perspectives.

%************************************************************************************
%************************************************************************************
\section{The setting}
\label{sec:problem}
%************************************************************************************
%************************************************************************************

%************************************************************************************
\subsection{Random matrix ensembles}
\label{rmt}
%************************************************************************************
 Let us first consider the case of a single Gaussian random matrix $H$ of size $N$, whose
   distribution is proportional to $\exp(-\frac12 \tr H^2)$. 
We denote by $\lambda_1\leq\ldots\leq\lambda_N$ the eigenvalues of such a matrix. The density of eigenvalues is given by the Wigner semicircle law  $\rho(\lambda)=\frac{1}{\pi N}\sqrt{2N-\lambda^2}$ \cite{wigner1957statistical}. Since there are $N\rho(\lambda)\delta \lambda$ levels in an interval $\delta \lambda$, the corresponding mean level spacing in the vicinity of $\lambda=0$ is $\Delta=\pi/\sqrt{2N}$, which gives a local density $1/\Delta=\sqrt{2N}/\pi$. The joint distribution of eigenvalues is~\cite{mehta2004random}
\begin{equation}
\label{joinedGOE}
P(\lambda_1,\ldots,\lambda_N)=\caln \prod_{i<j}(\lambda_j-\lambda_i)^\beta e^{-a\sum_{i=1}^N\lambda_i^2}\,,
\end{equation}
$\beta=1,2$ or 4 is the Dyson index and $\caln$ and $a$ are normalization constants. 

In a region of constant density, the nearest-neighbour spacing distribution $p(s)$ is well-approximated by the Wigner surmise~\cite{wigner_1955}, corresponding to the exact result obtained from Eq.~\eqref{joinedGOE} for $2\times 2$ matrices,
\begin{equation}
\label{wignersurmize}
p(s)=a_\beta s^\beta e^{-b_\beta s^2},
\end{equation}
where $a_\beta,b_\beta$ are normalization constants, chosen in such a way that $\langle s\rangle=1$. In a similar way \cite{atas2013distribution}, one can approximate the joint distribution of consecutive nearest-neighbour spacings $p(s,t)$ by its exact expression for $3\times 3$ matrices, which can be obtained from Eq.~\eqref{joinedGOE} with $N=3$ by integrating over one variable. It reads
\begin{equation}
\label{wignersurmize3}
p(s,t)=A_\beta s^\beta t^\beta (s+t)^\beta e^{ -B_\beta\left(s^2+s t+t^2\right)}
\end{equation}
where the constant $B_\beta$ is such that both spacings are normalized  as $\langle s\rangle=\langle t\rangle=1$, and $A_\beta$ is the overall normalization factor. From this expression, one can then obtain the distribution of $r=\min(t/s,s/t)$ as
\begin{align}
\label{pr2}
p(r)&=\int_0^{\infty}dsdt\;p(s,t)\delta\left(r-\min\left(\frac{s}{t},\frac{t}{s}\right)\right)\\
&=\int_0^{\infty}ds\; s\;\left(p(s,rs)+p(r s,s)\right).
\end{align}
Since the distribution $p(s,t)$ is symmetric in $s$ and $t$, Eq.~\eqref{pr2} reduces to
\begin{equation}
\label{pr2sym1bloc}
p(r)=2\int_0^{\infty}ds\; s\;p(s,rs).
\end{equation}
This approach was carried out in \cite{atas2013distribution}, yielding 
\begin{equation}
\label{prRMT}
p(r)=\frac{1}{Z_{\beta}}\frac{(r+r^2)^{\beta}}{(1+r+r^2)^{1+\frac{3}{2}\beta}},
\end{equation}
with $Z_{\beta}$ the normalization constant. 

Since Gaussian and circular ensembles have the same asymptotic level spacing distribution, the same analysis should equally be valid for circular ensembles, the only difference being that the mean level spacing is $\Delta=2\pi/N$ and thus the density of states is uniform and proportional to $N$, rather than circular and proportional to $\sqrt{N}$. For finite $N$, this difference in the shape of the density can result in small differences between the circular and Gaussian ensembles which are expected to vanish in the large-matrix limit. For the $3 \times 3$ matrices leading to the surmise Eq.~\eqref{prRMT}, the difference is already very small~\cite{dalessio_coe_2014}.

\subsection{Compound spectrum: the Rosenzweig-Porter approach}
Let us now consider ensembles of random matrices of size $N$ which can be decomposed into $m$ independent blocks of sizes $N_1,N_2 \ldots N_m$, with $\sum_i N_i=N$. The ordered eigenvalues $\lambda_1<\lambda_2<\cdots <\lambda_N$ of such a matrix can be obtained by diagonalizing each block separately and ordering the eigenvalues, so that the spectra of the blocks are interlaced.
Let $\mathbf{N}=(N_1,N_2 \ldots N_m)$ be a vector of block sizes. The compound spectrum $\{\lambda_i,1\leq i\leq N\}$ can be characterized by its spacing distribution $P_{\mathbf{N}}(s)$, which is the distribution of gaps $s_i=\lambda_{i+1}-\lambda_{i}$. It can also be characterized by the gap ratio distribution $P_{\mathbf{N}}(\tilde{r})$, with $\tilde{r}_i=s_i/s_{i-1}$, or~\footnote{The reader should not get confused by the fact that we inverted the notations between $r$ and $\tilde{r}$ with respect to Refs.~\cite{atas2013distribution,atas2013joint}.} by the gap ratio distribution $P_{\mathbf{N}}(r)$, with $r_i=\min(s_i/s_{i-1},s_{i-1}/s_i) \in [0,1]$.

If there is a statistical symmetry between left and right intervals then the relation $P_{\mathbf{N}}(\tilde{r})=\frac{1}{\tilde{r}^2}P_{\mathbf{N}}(\frac{1}{\tilde{r}})$ holds, which entails that $P_{\mathbf{N}}(r)=2 P_{\mathbf{N}}(\tilde{r})\theta(1-\tilde{r})$~\cite{atas2013distribution}. In that case, the distributions of $\tilde{r}$ and $r$ essentially contain the same information. As we shall see, this is the case for the distributions considered in this paper, and therefore, as is often done in numerical simulations, we concentrate on the distribution $P_{\mathbf{N}}(r)$ with $r\in [0,1]$.

If the $m$ blocks are independent Gaussian random matrices given by the Wigner-Dyson ensembles with index $\beta$, then the spectrum of block $i$, $\{\lambda^{(i)}_q, 1\leq q\leq N_i\}$, is characterized by its mean level spacing around $\lambda=0$, given by $\Delta_i=\pi/\sqrt{2N_i}$, or by its local density $\rho_i=\sqrt{2N_i}/\pi$. The resulting spectrum obtained by the superposition of the $m$ spectra has density $\rho=\sum_i\rho_i$. Introducing the normalized densities $\mu_i=\rho_i/\rho$, we have $\mu_i=\sqrt{N_i/N}$. For the circular ensembles, where densities are uniform over the unit circle, $\mu_i=N_i/N$.

The Rosenzweig-Porter approach, which gives the nearest-neighbour spacing distribution $P(x)$ associated with the compound spectrum, consists in assuming that the compound spectrum is a superposition of independent and identically distributed spectra  with uniform density $\rho_i$ and with nearest-neighbour spacing distribution given by the surmise Eq.~\eqref{wignersurmize}. The computation proceeds by identifying that a gap in the compound spectrum can originate either from a gap in one of the spectra or from a gap between eigenvalues from two distinct spectra. Considering all possibilities and the probabilities attached to them leads to the spacing distribution $P(x)$. This approach is detailed in Sec.~\ref{pdes}. These results were extended in \cite{Berry_1984} by considering a mixed phase space, which amounts to adding Poisson blocks to a chaotic Hamiltonian. The work~\cite{Berry_1984} derives an explicit formula for $P(x)$ for a Poisson block and $(N-1)$ chaotic blocks with same density.

\subsection{Summary of our results}
\label{secsummary}
In this paper we extend the Rosenzweig-Porter approach to derive the joint distribution of consecutive nearest-neighbour spacings $P(x,y)$ of a compound spectrum made out of several spectra with arbitrary distribution. It is given by the very compact expression Eq.~\eqref{pxy}--\eqref{Hxyfinal}, for which we give a probabilistic interpretation. We then obtain $P(r)$ from the analog of Eq.~\eqref{pr2sym1bloc}, namely
\begin{equation}
\label{pr2sym}
P(r)=2\int_0^{\infty}dx\; x\;P(x,r x).
\end{equation}
Applying our expressions to the RMT expressions Eqs.~\eqref{wignersurmize} and~\eqref{wignersurmize3}, we obtain a closed general expression for $P_{\mathbf{N}}(r)$. We then apply this general formula to the case of identical block sizes $N_i=N/m$, for which we use the short notation $P_m(r)$. Some of these calculations result in exact closed (albeit complex) forms, others require a numerical integration. Besides the full distribution, we will also consider the average gap ratio $\langle r \rangle_{m}=\int_0^1 r P_m(r) dr$ and the limiting value for vanishing gap ratio $P_m(0)=\lim_{r\rightarrow 0}P_m(r)$, as they turn out to be of great practical use to identify the existence of a symmetry ($P(0)=0$ in the no-symmetry case $m=1$). In Section \ref{sec:smallsize} we also consider the quantity $I_m^{1/4}=\int_0^{1/4}dr\,P(r)$, which proves useful to identify symmetries in an experimental setting where few realizations of the spectrum are available. Our results are summarised in Table \ref{tabcoef}. In the electronic supplementary material, we provide a Mathematica notebook allowing to reproduce our calculations.

\begin{table}[ht]
\centering
\begin{tabular}{| c | c | c | c | c | c |}
\hline
  & $m$ & GOE      &   GUE & GSE  \\
  \hline
  $\langle r \rangle$ ~\cite{atas2013distribution} & 1 & 0.53590 & 0.60266 & 0.67617 \\
\hline
 \multirow{13}{*}{$\langle r\rangle_m$}& 2  & 0.423415 & 0.422085 & 0.411762 \\
 & 3 &  0.403322 & 0.399229 & 0.392786\\
 & 4 &  0.396125 & 0.39253 & 0.388686\\
  & 5 &  0.392712 & 0.389805 & 0.387367 \\
  & 6 &  0.390821 & 0.388475 & 0.38684 \\
 & 7 &  0.389661 & 0.387745  & 0.386597 \\
   & 8 &  0.388898 & 0.387309 & 0.386474 \\
  & 9 &  0.388368 & 0.387033 & 0.386407 \\
  & 10 &  0.387986 & 0.386849 & 0.386368 \\
  & 11 &  0.387701 & 0.386721  & 0.386344 \\
  & 12 &  0.387482 & 0.38663 & 0.386329 \\
 & $\dots$ & $\dots$ &  $\dots$ &  $\dots$\\
 \cline{3-5}
 & $\infty$ (Poisson)  & \multicolumn{3}{c|}{0.386294} \\
\hline
\hline
 \multirow{13}{*}{$P_m(0)$} & 2 &  1.40805 & 1.5228 & 1.63484 \\
& 3 &  1.71587 & 1.80758 & 1.88322\\
 & 4 &  1.83279 & 1.9023 & 1.95178\\
   & 5 & 1.88972 & 1.94334 & 1.97682  \\
  & 6 & 1.92175 & 1.96413 & 1.98765  \\
  & 7 & 1.94157 & 1.97582 & 1.9929   \\
   & 8 & 1.95469 & 1.98292 & 1.99568  \\
  & 9 & 1.96383 & 1.98748 & 1.99724  \\
  & 10 & 1.97046 & 1.99055 & 1.99817  \\
  & 11 & 1.97541 & 1.99269 & 1.99875   \\
  & 12 & 1.97922 & 1.99423 & 1.99912  \\
 & $\dots$ &$\dots$ & $\dots$&$\dots$\\
 \cline{3-5}
  & $\infty$ (Poisson)  & \multicolumn{3}{c|}{2}\\
\hline
\end{tabular}
\caption{Values of averages $\langle{r}\rangle$ and probability at $r=0$ for $m$ blocks, obtained from the surmise approach in Sec.~\ref{sec:analytics1}. The value for $m=1$ is taken from \cite{atas2013distribution}. Values for $\langle r \rangle_{m}$ obtained from numerical simulations of random matrices are presented in Tab.~\ref{tab:rgaps} in Sec.~\ref{sec:numerics}. 
\label{tabcoef}}
\end{table}

%\clearpage

\section{Analytical results}
\label{sec:analytics1}

We now turn to the detailed proofs of our analytical formulae. The reader not interested in the details of the derivation can directly jump to Sec. III.C for a comparison to random matrix numerics, Sec. III.D for a discussion on how to compare to experimental results or to Sec. IV for several applications in many-body physics.

%************************************************************************************
\subsection{Nearest-neighbour spacing distribution $P(x)$}
\label{pdes}
%************************************************************************************
Since the function $p(s)$ in Eq.~\eqref{wignersurmize} corresponds to spectra with mean level spacing equal to 1, the spacing distributions of each spectrum are given by the function Eq.~\eqref{wignersurmize} rescaled by the mean level spacing, i.e.~$p(\rho_i s)$.
We introduce the functions
\begin{equation}
\label{defr}
\ff(s)=\int_0^\infty da\; p(s+a)
\end{equation}
and
\begin{equation}
\label{defd}
\fg(s)=\int_0^\infty da\; \int_0^\infty db\; p(s+a+b).
\end{equation}
The function $\ff(s)$ gives the probability to have $\lambda_{i+1}\geq s$ knowing that $\lambda_i=0$.
The function $\fg(s)$ gives the probability to have $\lambda_{i+1}\geq s$ knowing that $\lambda_i\leq 0$, that is, the probability to have a spacing at least $s$.
These probabilities are related through the identities
\begin{equation}
\label{deriv1}
\fg'=-\ff,\quad \fg''=p.
\end{equation}
Introducing the rescaled spacing $x=\rho s$ we have from Eq.~\eqref{deriv1}
\begin{align}
\label{rescalingps}
\fg(\rho_i s)&=\fg(\mu_i x),
\quad  \nonumber \\
\mu_i\ff(\rho_i s)&=-\partial_x \fg(\mu_i x),
\quad \nonumber \\
\mu_i^2 p(\rho_i s)&=\partial^2_x \fg(\mu_i x).
\end{align}
Spacings arise as empty intervals $]\lambda^{(i)}_q,\lambda^{(j)}_{q'}[$ of length $s$ with $i,j=1,...,m$. We have to consider the two possibilities $i=j$ or $i\neq j $, and calculate the probability densities associated with each configuration. 

Let us first consider the case where $m=2$ (we will later generalize this analysis to more blocks). We have to consider the two following cases:
\begin{itemize}
    \item[1.] Configurations giving rise to an empty interval of type $]\lambda^{(i)}_q,\lambda^{(i)}_{q+1}[$, which are are such that $\lambda^{(j)}_{q'}<\lambda^{(i)}_q<\lambda^{(i)}_{q+1}<\lambda^{(j)}_{q'+1}$ for some $q'$. The probability of such a configuration for $i$ is given by $p(\rho_i s)$, while the probability for $j$ is $\fg(\rho_j s)$ since $\lambda^{(j)}_{q'}$ and $\lambda^{(j)}_{q'+1}$ can be anywhere outside $]\lambda^{(i)}_q,\lambda^{(i)}_{q+1}[$. Taking into account the probability $\mu_i^2$ to have a level $i$ at both ends of the interval, we get for the configuration $]\lambda^{(i)}_q,\lambda^{(i)}_{q+1}[$ a probability density $\mu_i^2p(\rho_i s)\fg(\rho_j s)$. Using Eq.~\eqref{rescalingps} we can rewrite it as $[\partial^2_x \fg(\mu_i x)]\fg(\mu_j x)$.
    \item[2.] Configurations giving rise to an empty interval of type $]\lambda^{(i)}_q,\lambda^{(j)}_{q'}[$, which are such that  $\lambda^{(j)}_{q'-1}<\lambda^{(i)}_q<\lambda^{(j)}_{q'}<\lambda^{(i)}_{q+1}$. The probability of such a spacing for $i$ is given by $\ff(\rho_i s)$ since $\lambda^{(i)}_{q+1}$ can be anywhere in $]\lambda^{(j)}_{q'},\infty[$, while the probability for $j$ is $\ff(\rho_j s)$ since $\lambda^{(j)}_{q'-1}$ can be anywhere in $]-\infty,\lambda^{(i)}_q[$. The probability of having a level $i$ and a level $j$ at the ends of the interval is given by $\mu_i\mu_j$, so that the probability density of configuration $]\lambda^{(i)}_q,\lambda^{(j)}_{q'}[$ is $\mu_i\mu_j \ff(\rho_i s)\ff(\rho_j s)$. Using Eq.~\eqref{rescalingps} it can be rewritten as $\left[\partial_x \fg(\mu_i x)\right]\left[\partial_x \fg(\mu_j x)\right]$. 
\end{itemize}

Summing these probabilities over $i,j=1,2$ we get for the spacing probability $P(x)$ of the mixed levels
\begin{align}
P(x)&=\fg(\mu_1 x)\partial^2_x \fg(\mu_2 x)+\fg(\mu_2 x)\partial^2_x \fg(\mu_1 x)
\nonumber \\ 
&+2\left[\partial_x \fg(\mu_1 x)\right]\left[\partial_x \fg(\mu_2 x)\right]
%\nonumber \\ &
%=\partial^2_x \left(\fg(\mu_1 x)\fg(\mu_2 x)\right),
\end{align}
or equivalently
\begin{equation}
P(x)=\partial^2_x \fgg(x),\qquad \fgg(x)=\prod_{i=1}^m\fg(\mu_i x).
\label{pd}
\end{equation}

The above reasoning proves Eq.~\eqref{pd} for $m=2$. In fact, given the final expression, we can come up with a much shorter proof of the validity of Eq.~\eqref{pd} for arbitrary $m$. Indeed, the probability of finding an interval of a given length between two consecutive eigenvalues is the second derivative of the probability of finding an interval larger or equal to it with no eigenvalue in it (see Eqs.~\eqref{defd} and \eqref{deriv1}). Therefore $P(x)$ must be the second derivative of the probability of finding an empty interval larger or equal to $x$. The probability that no level of type $i$ occurs in an interval of size $x$ is $\fg(\mu_i x)$. Since levels from different sequences are independent, the probability that no level of any type occur in an interval of size $x$ is simply the product of all $\fg(\mu_i x)$, which directly entails Eq.~\eqref{pd}. Incidentally one can check, using Eqs.~\eqref{defr}--\eqref{deriv1}, that $P(x)$ in Eq.~\eqref{pd} is properly normalized to 1.

%************************************************************************************
\subsection{Joint consecutive spacing distribution $P(x,y)$}
%************************************************************************************
\label{sec:jointPxy}
We now apply the same line of reasoning to the joint distribution of two consecutive spacings. Our aim is to obtain the joint distribution $P(x,y)$ in terms of the distribution $p(s,t)$ for a single spectrum. 

Starting with the function $p(s,t)$ in Eq.~\eqref{wignersurmize3}, we introduce the function
\begin{equation}
\label{def2p}
\phat(s)=\int_0^\infty da\; p(s,a),
\end{equation}
which is the marginal distribution of $p(s,t)$. Since $p(s,t)$ defined in Eq.~\eqref{wignersurmize3} is symmetric in the exchange of $s$ and $t$, the marginal distribution can be equivalently taken by integrating over the first variable. Note that this expression differs from the one in Eq.~\eqref{wignersurmize}, which corresponds to the result for $2\times 2$ matrices while $p(s,t)$ was obtained for $3\times 3$ matrices. Functions $\ff$ and $\fg$ can then be defined from $\hat{p}$ as in Eq.~\eqref{defr}--\eqref{defd}. In terms of $p(s,t)$, their explicit form is 
\begin{equation}
\label{def2r}
\ff(s)=\int_0^\infty da\; \int_0^\infty db\; p(s+a,b)
\end{equation}
and
\begin{equation}
\label{def2d}
\fg(s)=\int_0^\infty da\; \int_0^\infty db\; \int_0^\infty dc\; p(s+a+b,c).
\end{equation}
We also define the two-variable functions
\begin{equation}
\label{def2e}
\fea(s,t)=\int_0^\infty da\, p(s+a,t), \quad \feb(s,t)=\int_0^\infty da\, p(s,t+a),
\end{equation}
and 
\begin{equation}
\label{def2s}
\fh(s,t)=\int_0^\infty da\; \int_0^\infty db\; p(s+a,t+b).
\end{equation}
These functions are related through the identities
\begin{equation}
\label{deriv2}
\partial_s\partial_t\fh=p,~ \partial_s\fh=-\feb,~ \partial_t\fh=-\fea, ~ g'=-f,~ g''=\phat.
\end{equation}
The analogs of Eqs.~\eqref{rescalingps} are  
\begin{align}
\label{rescalingpr1}
&\fg(\rho_i s)=\fg(\mu_i x),\quad \mu_i\ff(\rho_i s)=-\partial_x \fg(\mu_i x),\\
 &\mu_i^2 p(\rho_i s,\rho_i t)=\partial_x\partial_y \fh(\mu_i x,\mu_i y)\\
&\partial_x\fh(\mu_i x,\mu_i y)=-\mu_i \feb(\rho_i s,\rho_i t)\\
&\partial_y\fh(\mu_i x,\mu_i y)=-\mu_i \fea(\rho_i s,\rho_i t).
\label{rescalingpr2}
\end{align}

A sequence of two consecutive spacings arises as a sequence of two empty intervals $]\lambda^{(i)}_q,\lambda^{(j)}_{q'}[$ and $]\lambda^{(j)}_{q'},\lambda^{(k)}_{q''},[$ with $i,j,k=1,...,m$.  We have to consider all possibilities for $i,j,k$ and calculate the probability densities associated with each configuration. 

Once again, let us consider the simplest case $m=3$, that we will later generalize. We only need to examine four cases, corresponding to patterns $iii$, $iij$, $iji$ and $ijk$ and depicted in Fig.~\ref{fig:spacingsijk} :
\begin{figure}[!t]
\includegraphics[width=\columnwidth]{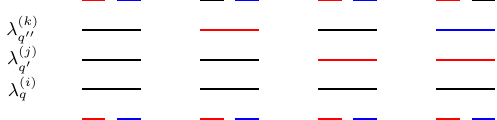}
\caption{
The four configurations of consecutive spacings considered in Sec.~\ref{sec:jointPxy}. Different colours correspond to distinct spectra. The three central levels are the ones from which the ratios are calculated; the outer levels are at the same height to indicate that their relative position is irrelevant.}
\label{fig:spacingsijk}
\end{figure}
\begin{itemize}
    \item[1.] Configurations $\lambda^{(i)}_q<\lambda^{(i)}_{q+1}<\lambda^{(i)}_{q+2}$, which arise whenever $\lambda^{(j)}_{q'}<\lambda^{(i)}_q<\lambda^{(i)}_{q+1}<\lambda^{(i)}_{q+2}<\lambda^{(j)}_{q'+1}$ for some $q'$ and $\lambda^{(k)}_{q''}<\lambda^{(i)}_q<\lambda^{(i)}_{q+1}<\lambda^{(i)}_{q+2}<\lambda^{(k)}_{q''+1}$ for some $q''$. The probability of such a configuration for $i$ is given by $p(\rho_i s, \rho_i t)$; the probability for $j$ is $\fg(\rho_j (s+t))$ since $\lambda^{(j)}_{q'}$ and $\lambda^{(j)}_{q'+1}$ can be anywhere outside $]\lambda^{(i)}_q,\lambda^{(i)}_{q+2}[$; and the same goes for $k$. Taking into account the probability $\mu_i^3$ to have three levels $i$ at the ends of the intervals, we get for this configuration a probability density $\mu_i^3p(\rho_i s,\rho_i t)\fg(\rho_j (s+t))\fg(\rho_k (s+t))$. Using Eqs.~\eqref{rescalingpr1}--\eqref{rescalingpr2} we can rewrite it as $\mu_i[\partial_x\partial_y \fh(\mu_i x,\mu_i y)]\fg(\mu_j (x+y))\fg(\mu_k (x+y))$.
    \item[2.] Configurations $\lambda^{(i)}_q<\lambda^{(i)}_{q+1}<\lambda^{(j)}_{q'}$, which arise whenever $\lambda^{(j)}_{q'-1}<\lambda^{(i)}_q<\lambda^{(i)}_{q+1}<\lambda^{(j)}_{q'}<\lambda^{(i)}_{q+2}$ and  $\lambda^{(k)}_{q''}<\lambda^{(i)}_q<\lambda^{(i)}_{q+1}<\lambda^{(j)}_{q'}<\lambda^{(k)}_{q''+1}$ for some $q''$. The probability for $i$ is $\feb(\rho_i s, \rho_i t)$, the probability for $j$ is $\ff(\rho_j (s+t))$, while the probability for $k$ is $\fg(\rho_k (s+t))$. We get for this configuration a probability density $\mu_i^2\mu_j \feb(\rho_i s, \rho_i t)\ff(\rho_j (s+t))\fg(\rho_k (s+t))=\mu_i[\partial_x\fh(\mu_i x,\mu_i y)][\partial_y\fg(\mu_j (x+y)]\fg(\mu_k (x+y))$. We used the fact that $\partial_y\fg(x+y)=\fg'(x+y)$.
    \item[3.] Configurations $\lambda^{(i)}_q<\lambda^{(j)}_{q'}<\lambda^{(i)}_{q+1}$ which arise whenever $\lambda^{(j)}_{q'-1}<\lambda^{(i)}_q<\lambda^{(j)}_{q'}<\lambda^{(i)}_{q+1}<\lambda^{(j)}_{q'+1}$ and  $\lambda^{(k)}_{q''}<\lambda^{(i)}_q<\lambda^{(j)}_{q'}<\lambda^{(i)}_{q+1}<\lambda^{(k)}_{q''+1}$ for some $q''$. The probability for $i$ is $\phat(\rho_i (s+t))$, the probability for $j$ is $\fh(\rho_j s,\rho_j t)$, and the probability for $k$ is $\fg(\rho_k (s+t))$. We get for this configuration a probability density $\mu_i^2\mu_j \phat(\rho_i (s+t))\fh(\rho_j s,\rho_j t)\fg(\rho_k (s+t))=\mu_j[\partial_x\partial_y\fg(\mu_i(x+y))]\fh(\mu_j x, \mu_j y)\fg(\mu_k (x+y))$. We used the fact that $\partial_x\partial_y\fg(x+y)=\fg''(x+y)$.
    \item[4.] Finally, configurations $\lambda^{(i)}_q<\lambda^{(j)}_{q'}<\lambda^{(k)}_{q''}$, which arise whenever $\lambda^{(j)}_{q'-1}<\lambda^{(i)}_q<\lambda^{(j)}_{q'}<\lambda^{(k)}_{q''}<\lambda^{(j)}_{q'+1}$ and $\lambda^{(k)}_{q''-1}<\lambda^{(i)}_q<\lambda^{(j)}_{q'}<\lambda^{(k)}_{q''}<\lambda^{(i)}_{q+1}$. The probability for $i$ is $\ff(\rho_i (s+t))$, the probability for $j$ is $\fh(\rho_j s,\rho_j t)$, and the probability for $k$ is $\ff(\rho_k (s+t))$. We get for this configuration a probability density $\mu_i\mu_j\mu_k \ff(\rho_i (s+t))\fh(\rho_j s,\rho_j t)\ff(\rho_k (s+t))=\mu_j[\partial_x\fg(\mu_i (x+y))]\fh(\mu_j x, \mu_j y)[\partial_y\fg(\mu_k (x+y))]$. 
\end{itemize}

We can now sum all contributions over $i,j,k=1,2,3$. There are 27 terms, which can be put under the compact form 
\begin{align}
P(x,y)&=\partial_x\partial_y[\mu_1\fh(\mu_1 x, \mu_1 y)\fg(\mu_2 (x+y))\fg(\mu_3 (x+y))\nonumber\\
&+\mu_2\fg(\mu_1 (x+y))\fh(\mu_2 x, \mu_2 y)\fg(\mu_3 (x+y))\nonumber\\
&+\mu_3\fg(\mu_1 (x+y))\fg(\mu_2 (x+y))\fh(\mu_3 x, \mu_3 y)].
\label{defH}
\end{align}
Equation \eqref{defH} has a simple probabilistic interpretation. Let us define a function $\fhh$ as 
\begin{equation}
\label{def2sbis}
\fhh(x,y)=\int_0^\infty da\; \int_0^\infty db\; P(x+a,y+b),
\end{equation}
by analogy with Eq.~\eqref{def2s}. Thus $\fhh(x,y)$ gives the probability of having a triplet $(\lambda_{q-1},\lambda_{q},\lambda_{q+1})$ of levels of the mixed spectrum such that $\lambda_{q-1}<\lambda_{q}-x$ and $\lambda_{q}+y<\lambda_{q+1}$. That is, $\fhh(x,y)$ is the probability that some level 
 $\lambda^{(i)}_{q}$ is such that all other levels $\lambda^{(j)}_{q'}$ verify either $\lambda^{(j)}_{q'}<\lambda^{(i)}_{q}-x$ or $\lambda^{(i)}_{q}+x<\lambda^{(j)}_{q'}$ (including the case $i=j$, in which case of course $q'\neq q$). At fixed $i$, the probability of such a configuration is $\fh(\mu_i x, \mu_i y)$ for spectrum $i$, and $\fg(\mu_j (x+y))$ for all $j\neq i$. Summing over all $i$ (and taking into account the probability $\mu_i$ to have a level $i$ in the middle), we get for $H(x,y)$ the expression under the derivation symbols in Eq.~\eqref{defH}. 
 
 In fact, this reasoning provides a proof of the general case with arbitrary number $m$ of spectra. We thus have in the general case
\begin{equation}
\label{pxy}
P(x,y)=\partial_x\partial_y \fhh(x,y)
\end{equation}
with 
\begin{equation}
\label{Hxyfinal}
H(x,y)=\sum_{i=1}^m\mu_i\fh(\mu_i x, \mu_i y)\prod_{j\neq i}\fg(\mu_j (x+y)).
\end{equation}
One can check, using Eqs.~\eqref{deriv2}--\eqref{rescalingpr2}, that $P(x,y)$ in Eq.~\eqref{pxy} is properly normalized to 1. Note that, although in what follows we will apply Eqs.~\eqref{pxy}--\eqref{Hxyfinal} to the random matrix case, these equations are valid for an arbitrary initial distribution $p(s,t)$ of individual spectra.

%************************************************************************************
\subsection{Ratio distribution $P(r)$}
\label{subsec:ratio_surmise}
%************************************************************************************

In order to obtain $P(r)$, one first needs to evaluate functions $g$ and $h$ to obtain $H(x,y)$ using Eq.~\eqref{Hxyfinal}, then take its derivative with respect to $x$ and $y$, and finally perform the integration in Eq.~\eqref{pr2sym}. In the GUE and GSE case, there is no closed-form expressions for the function $h$, so that we are left with a double integral (one in the definition of $h$, one corresponding to the final integration in Eq.~\eqref{pr2sym}).
In the GOE case however, we obtain explicit expressions for $g$ and $h$, as we will show below, and thus we get a closed-form expression for $P(x,y)$. The remaining integral Eq.~\eqref{pr2sym} is doable analytically only in the case of a mixture of $m=2$ spectra.

\subsubsection{GOE case}
In the GOE case, the joint distribution $p(s,t)$ reads
\begin{equation}
\label{wignersurmizeGOE}
p(s,t)=\frac{3^7}{2^5 \pi ^3} s t (s+t) e^{ -\frac{9}{4 \pi}\left(s^2+s t+t^2\right)}.
\end{equation}
Starting from Eq.~\eqref{wignersurmizeGOE} for $p(s,t)$ and calculating explicitly the functions $\fg$ and $\fh$ given in Eqs.~\eqref{def2d} and \eqref{def2s} we get
\begin{equation}
\label{gU}
\fg(s)=U_1(s)-\frac{s}{2}U_2(s)-\frac{s}{2}U_3(s)
\end{equation}
with 
\begin{align}
U_1(s)&=e^{ -\frac{9}{4 \pi}s^2},
\nonumber \\
U_2(s)&=\erfc\left(\frac{3s}{2\sqrt{\pi}}\right), 
\nonumber \\ 
U_3(s)&=e^{ -\frac{27}{16 \pi}s^2}\erfc\left(\frac{3s}{4\sqrt{\pi}}\right),
\end{align}
and
\begin{align}
\label{hV}
\fh(s,t)&= 
\frac{9(s+t)}{4\pi}V_1(s,t)
\nonumber \\
&+\frac{8\pi-27s^2}{16 \pi}V_2(s,t)+\frac{8\pi-27t^2}{16 \pi}V_3(s,t)
\end{align}
with 
\begin{align}
V_1(s,t)&=e^{ -\frac{9}{4 \pi}(s^2+s t+t^2)}, \nonumber \\
V_2(s,t)&=e^{ -\frac{27s^2}{16 \pi}}\erfc\left(\frac{3(s+2t)}{4\sqrt{\pi}}\right),  \nonumber \\
V_3(s,t)&=e^{ -\frac{27t^2}{16 \pi}}\erfc\left(\frac{3(2s+t)}{4\sqrt{\pi}}\right).
\end{align}
One can then rewrite Eq.~\eqref{Hxyfinal} as
%\begin{widetext}
\begin{align}
\label{Hxyrew}
& H(x,y)=
\nonumber \\
& \sum_{i=1}^m \sum_{a_1=1}^3\cdots\sum_{a_m=1}^3 
 H^{(i)}_{a_1\ldots a_m}V_{a_i}(\mu_i x, \mu_i y)  \prod_{j\neq i}U_{a_j}(\mu_j (x+y)) 
\end{align}
%\end{widetext}
with $H^{(i)}_{a_1\ldots a_m}$ some polynomials of $x$, $y$ and the $\mu_i$, which can be obtained explicitly from Eqs.~\eqref{gU} and~\eqref{hV}. 

Functions $U_i$ have the property that they transform into each other under derivation. Namely, the derivative of any function $\sum_i c_iU_i$ (with $c_i$ polynomial in $s$) is of the form $\sum_i \tilde{c}_iU_i$ (with $\tilde{c}_i$ polynomial in $s$). The same property holds for the $V_i$ upon derivation with respect to $s$ or $t$. Therefore, using Eq.~\eqref{pxy} and the expansion Eq.~\eqref{Hxyrew}, we obtain 
%\begin{widetext}
\begin{align}
\label{Pxyrew}
& P(x,y)= \nonumber \\
& \sum_{i=1}^m \sum_{a_1=1}^3\cdots\sum_{a_m=1}^3 P^{(i)}_{a_1\ldots a_m}V_{a_i}(\mu_i x, \mu_i y)\prod_{j\neq i}U_{a_j}(\mu_j (x+y) )
\end{align}
%\end{widetext}
 where $P^{(i)}_{a_1\ldots a_m}$ are polynomials of $x$, $y$ and the $\mu_i$. Given the definition of the functions $U_i$ and $V_i$, Eq.~\eqref{pr2sym} can be expanded as a linear combination (with real coefficients dependent on the $\mu_i$) of integrals of the form
\begin{equation}
\label{intmystere}
\int_0^\infty dx\;x^k e^{-\lambda x^2}\prod_{i=1}^m\erfc(a_i x),
\end{equation}
with $\lambda$ and the $a_i$ depending on the $\mu_i$ and on $r$ (and possibly $a_i=0$).
It appears that in general such an integral does not have a closed form. However in the case $m=2$ we have the identity
\begin{align}
\int_0^\infty dx\;x e^{-\lambda x^2}\erfc(u x)\erfc(vx)=
\nonumber\\
\frac{1}{2 \lambda }-\frac{u \tan ^{-1}\left(\frac{\sqrt{\lambda +u^2}}{v}\right)}{\pi  \lambda  \sqrt{\lambda
   +u^2}}-\frac{v \tan ^{-1}\left(\frac{\sqrt{\lambda +v^2}}{u}\right)}{\pi  \lambda  \sqrt{\lambda +v^2}},
\end{align}
from which one can deduce Eq.~\eqref{intmystere} for all odd values of $k$ by deriving with respect to $\lambda$, and for all even values of $k$ by first integrating by parts and then deriving with respect to $\lambda$. This yields a (rather lengthy) closed-form expression for $P(r)$ in the case $m=2$ (which is given in full in the electronic Supplementary Material). To give an idea of this expression, we only give $P(0)$ in the case of two blocks of the same size:
\begin{align}
    P(0)&=\frac{1}{168} \left(408-144 \sqrt{2}+7 \sqrt{6} \pi \right.
    \nonumber \\
    &  \left. +14 \sqrt{6} \tan ^{-1}\left(\frac{1}{4 \sqrt{3}}\right)-28 \sqrt{6} \tan
   ^{-1}\left(\frac{1}{\sqrt{6}}\right)\right)
   \nonumber \\
 &  \simeq 1.40805.
\end{align}
For $m$ blocks of the same size, we get $P(0)=1.71587$ for $m=3$ and $P(0)=1.83279$ for $m=4$, as reported in Tab.~\ref{tabcoef}.

In practice, the fastest way of obtaining $P(r)$ for GOE in the general case is to calculate $H(x,y)$ and $P(x,y)$ analytically from the explicit expressions for $h$ and $g$, using Eq.~\eqref{pxy}--\eqref{Hxyfinal}, and perform the last integral numerically. The Mathematica notebook in the electronic Supplementary Material implements the two possibilities to obtain $P(r)$. From the exact equation, numerical integration over $[0,1]$ yields the mean ratio. For instance for $m$ blocks of equal size, we get 
\begin{equation}
\langle r\rangle_{\textrm{GOE}, m \textrm{ blocks}}=0.423415,\quad 0.403322,\quad 0.396125
\end{equation}
for $m=2,3$ and 4 blocks respectively, again reported in Tab.~\ref{tabcoef}.

\subsubsection{GUE case}
In the GUE case, the joint distribution $p(s,t)$ reads
\begin{equation}
\label{wignersurmizeGUE}
p(s,t)=\frac{3^{23}\sqrt{3}}{2^{26} \pi^5} s^2 t^2 (s+t)^2 e^{ -\frac{243}{64\pi}\left(s^2+s t+t^2\right)}.
\end{equation}
The calculation of $g$ and $h$ for GOE was made possible by the fact that either $s$ or $t$ is of degree 1 in the polynomial in front of the exponential in Eq.~\eqref{wignersurmizeGOE}. This is no longer the case for GUE and GSE. However, $g$ as well as the first derivative of $h$ can be obtained analytically. We find
\begin{align}
    g(x)&=
   -\frac{729 \sqrt{3}x^3}{1024 \pi ^2}e^{-\frac{243 x^2}{64 \pi }}-\erfc\left(\frac{9}{8} \sqrt{\frac{3}{\pi }} x\right)
   \nonumber \\
   +   e^{-\frac{729 x^2}{256 \pi }} & \left(\frac{243 \left(243
   x^4+128 \pi  x^2\right)}{49152 \pi ^2}+2\right)
  \erfc\left(\frac{9}{16} \sqrt{\frac{3}{\pi }} x\right)\nonumber\\
   & -\frac{3}{2} x \left(\erfc\left(\frac{27 x}{16 \sqrt{\pi }}\right)-4 T\left(\frac{27 x}{8 \sqrt{2 \pi
   }},\frac{1}{\sqrt{3}}\right)\right)
   \label{gGUE}
\end{align}
where $T(x,a)$ is Owen's $T$-function, defined as
\begin{equation}
    T(x,a)=\frac{1}{2\pi}\int_{0}^a 
dt\frac{1}{1+t^2}e^{-x^2(1+t^2)/2}.
\end{equation}
For $h$ we have $\fh(x,y)=\int_{0}^{\infty}\feb(x+a,y)da$, with
\begin{align}
 e_2(x,y)&=\int_0^\infty db\; p(x,y+b)
 \nonumber \\
 &=\frac{3^{14} \sqrt{3} x^2 e^{-\frac{243 \left(x^2+x y+y^2\right)}{64 \pi }} (x+2 y)    }{2^{24}
   \pi ^4}\nonumber\\
  & \times \left(128 \pi -81 \left(x^2-4 x y-4 y^2\right)\right) 
  \nonumber \\  
& 
+
\frac{3^{12} e^{-\frac{729 x^2}{256 \pi }} x^2 \erfc\left(\frac{9}{16} \sqrt{\frac{3}{\pi }}
   (x+2 y)\right)}{2^{28} \pi ^4}
   \nonumber \\
   & \times \left(19683 x^4-20736 \pi  x^2+16384 \pi ^2\right) 
   \end{align}
and $\fea(x,y)=\feb(y,x)$. Using the identities in Eq.~\eqref{deriv2}, the expression of $P(x,y)$ reads
\begin{align}
P(x,y)&=\sum_{i=1}^m\mu_i\fh(\mu_i x, \mu_i y)\partial_x\partial_y\left[\prod_{j\neq i}\fg(\mu_j (x+y))\right]
\nonumber \\
& -
\sum_{i=1}^m\mu_i^2\feb(\mu_i x, \mu_i y)\partial_y\left[\prod_{j\neq i}\fg(\mu_j (x+y))\right]\nonumber\\
&-\sum_{i=1}^m\mu_i^2\fea(\mu_i x, \mu_i y)\partial_x\left[\prod_{j\neq i}\fg(\mu_j (x+y))\right]
\nonumber \\
&+
\sum_{i=1}^m\mu_i^3p(\mu_i x, \mu_i y)\prod_{j\neq i}\fg(\mu_j (x+y)),
\label{pxyGUE}
\end{align}
in which only the first sum involves the function $h$ for which no closed form is available. The $i$th term in that sum is $\mu_i\fh(\mu_i x, \mu_i y)q_i(x+y)$, with $q_i$ an explicitly known function involving only products of derivatives of $g$. 
For this term the integral Eq.~\eqref{pr2sym} yields a contribution 
\begin{align}
\label{twofold}
&2\mu_i\int_0^\infty dx \;x\; \fh(\mu_i x, \mu_i r x)\;q_i((1+r)x)
\nonumber \\
&=2\mu_i \int_0^\infty dx\;\int_{0}^{\infty}d a\; x\;\feb(\mu_i x+a,\mu_i r x) q_i((1+r)x).
\end{align}
We perform numerically the twofold integrals Eq.~\eqref{twofold} and the single integrals over $x$ for all the other terms.
For $m$ blocks of equal size, we obtain:
\begin{equation}
    P(0)=1.5228,\quad 1.80758, \quad 1.9023
\end{equation}
and
\begin{equation}
\langle r\rangle_{\textrm{GUE}, m \textrm{ blocks}}=0.422085,\quad 0.399229,\quad  0.39253
\end{equation}
for $m=2,3,4$ respectively.

\subsubsection{GSE case}
For GSE the joint distribution $p(s,t)$ reads
\begin{equation}
\label{wignersurmizeGSE}
p(s,t)=\frac{3^{81}\sqrt{3}}{2^{75}5^{15} \pi^8} s^4 t^4 (s+t)^4 e^{ -\frac{3^{11}}{2^{10}5^2\pi}\left(s^2+s t+t^2\right)}.
\end{equation}
The function $g$ reads
\begin{align}
    g(x)&=
   -\frac{3}{2} x \left(\erfc\left(\frac{729 x}{320 \sqrt{\pi }}\right)-4 T\left(\frac{729 x}{160 \sqrt{2 \pi
   }},\frac{1}{\sqrt{3}}\right)\right)
   \nonumber \\
     &  -\erfc\left(\frac{243}{160} \sqrt{\frac{3}{\pi }} x\right)
   \nonumber\\
&  +\frac{e^{-\frac{531441 x^2}{102400
   \pi }}  \erfc\left(\frac{243}{320} \sqrt{\frac{3}{\pi }} x\right)}{10368}Q_1(x)\nonumber\\
   &
  -\frac{531441
   \sqrt{3} e^{-\frac{177147 x^2}{25600 \pi }}    x^3}{262144000 \pi ^2} Q_2(x)
   \label{gGSE}
\end{align}
with $Q_1$, $Q_2$ the polynomials
\begin{align}
    Q_1(x)&=\frac{523347633027360537213511521 x^{10}}{10995116277760000000000 \pi ^5}
    \nonumber \\
    &+\frac{8862938119652501095929
   x^8}{214748364800000000 \pi ^4}
   \nonumber \\
   &+\frac{16677181699666569 x^6}{167772160000 \pi ^3}\nonumber\\
   &+\frac{2792914305201 x^4}{81920000 \pi
   ^2}+\frac{14703201 x^2}{400 \pi }+20736
\end{align}
and 
\begin{align}
Q_2(x) &= \frac{16677181699666569 x^6}{16777216000000 \pi ^3}
\nonumber \\ 
&+\frac{94143178827 x^4}{163840000 \pi ^2}+\frac{531441 x^2}{256 \pi }-32.
\end{align}
It can be checked that the second derivative of $g$ is indeed the marginal probability $\hat{p}$, and that $g(0)=1$. Similarly as in the GUE case, $h$ is not calculable in closed form, but we have
$\fh(x,y)=\int_{0}^{\infty}\feb(x+a,y)da$ with
\begin{align}
 e_2(x,y)&=\frac{3^{33} x^4}{2^{79}5^{14} \pi ^7}
 e^{-\frac{177147 \left(x^2+x y+y^2\right)}{25600 \pi }} \Bigg( R_1(x,y) 
 \nonumber\\
 & \left. +R_2(x,y) e^{\frac{177147 (x+2 y)^2}{102400 \pi }}\erf\left(\frac{243}{320} \sqrt{\frac{3}{\pi }} (x+2
   y)\right)\right),
 \end{align}
where $R_1(x,y)$ and $R_2(x,y)$ are polynomials given by
\begin{align}
    &R_1(x,y) =77760 \sqrt{3} (x+2 y) \times
    \nonumber \\
    & \left(\phantom{\frac12}\!\!\!\!-773967052800000 \pi ^2 \left(5 x^2-28 x y-28 y^2\right) +
    \right.\nonumber\\
   & 535570083993600 \pi  \left(5 x^4-24 x^3 y+88 x^2
   y^2+224 x y^3+112 y^4\right)\nonumber\\
    &-1853020188851841 \left(x^6-4 x^5 y+12 x^4 y^2-32 x^3 y^3 \right. \nonumber \\
    & \left.
    -176 x^2 y^4-192 x y^5-64 y^6\right)
%    \nonumber\\ &
    \left.+4697620480000000 \pi^3\phantom{\frac12}\!\!\!\!\right)
\end{align}
and
\begin{align}
    R_2(x,y)&=328256967394537077627 x^8
    \nonumber \\   &
   -379498534676857036800 \pi  x^6
      \nonumber \\   &
   +493581389408501760000 \pi ^2 x^4\nonumber\\
    &-475525357240320000000 \pi ^3   x^2
       \nonumber \\   &
   +240518168576000000000 \pi ^4.
    \end{align}
The computation is then the same as for GUE. Using Eqs.~\eqref{pxyGUE} and~\eqref{twofold} we get, for $m=2,3,4$ blocks of equal size,
\begin{equation}
    P(0)=1.63484,\quad 1.88322,\quad 1.95178
\end{equation}
and
\begin{equation}
\langle r\rangle_{\textrm{GSE}, m \textrm{ blocks}}=0.411762,\quad 0.392786,\quad 0.388686.
\end{equation}

We finally mention that the Mathematica notebook in the electronic Supplementary Material allows to reproduce these computations, as well as to consider different cases (other values of $m$, unequal sizes of the $m$ blocks).

\subsubsection{Poisson ($m\to\infty$) limit}
One can easily check that in the case of a mixture of $m$ spectra of the same size, one recovers the Poisson distribution in the $m\to\infty$ limit. Indeed, in that case the function $H$ reads
\begin{equation}
    H(x,y)=h\left(\frac{x}{m},\frac{y}{m}\right)g\left(\frac{(1+r)x}{m}\right)^{m-1}
\end{equation}
and thus
\begin{align}
    P(r)=2\int_0^\infty dx\;x\; & g^{m-1}
\left[\frac{\partial_x\partial_y h}{m^2}+\frac{m-1}{m^2}(\partial_xh+\partial_y h)\frac{g'}{g} \right.
    \nonumber\\
 & \left.  +h\frac{m-1}{m^2}\left((m-2)\frac{g'^2}{g^2}+\frac{g''}{g}\right)\right],
\end{align}
with functions $h$ and $g$ evaluated at $\left(\frac{x}{m},\frac{r x}{m}\right)$ and $\frac{(1+r)x}{m}$, respectively. In the limit $m\to\infty$, these arguments go to 0. From the explicit expressions for $g$ and $h$, the only term in the square brackets that survives is $h\frac{(m-1)(m-2)}{m^2}\frac{g'^2}{g^2}$, which goes to 1. Using the fact that $g(x)=1-x+O(x^3)$ close to 0, we get
\begin{align}
    P(r)&\simeq 2\int_0^\infty dx\;x\; \left(1-\frac{(1+r)x}{m}\right)^{m-1}
    \nonumber\\
    &\to_{m\to\infty}2\int_0^\infty dx\;x\;e^{-(1+r)x}=\frac{2}{(1+r)^2},
\end{align}
which is indeed the Poisson result.

\subsection{Comparison with numerics}
\label{sec:numerics}

%%%%%%%%%%%%%%%%%%%%%
\begin{figure}[htp]
\includegraphics[width=\columnwidth]{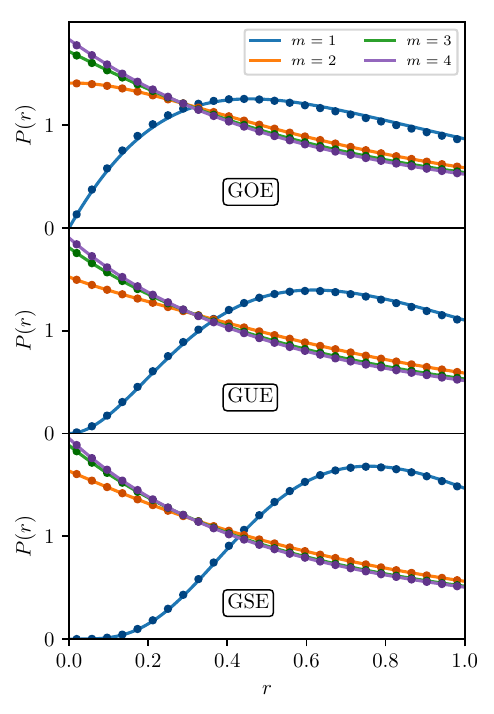}
\caption{Distribution of the ratio of consecutive level spacings $P(r)$ for (from top to bottom) GOE, GUE and GSE ensembles and $m=1$ to $m=4$ blocks (from bottom to top at $r=0$). Full lines are the
surmises obtained from Sec.\ \ref{sec:analytics1}, except the $m=1$ case, for which the corresponding surmise Eq.~\eqref{prRMT} is taken from \cite{atas2013distribution}. Points are numerical results from random matrices of size at least $N_\text{tot} = 2000$, averaged over $3.6\times10^5$ histograms.}
\label{fig:rmt}
\end{figure}
%%%%%%%%%%%%%%%%%%%%

We now compare the gap ratio distribution $P_m(r)$ and the mean value $\langle r \rangle_m$ obtained through the analytical approach of Sec.~\ref{sec:analytics1} to direct numerical computations on large random matrices.  
Numerical RMT spectra are computed using the matrix models of \cite{dumitriu_matrix_models}, based on
tridiagonal matrices. These models are numerically faster to diagonalize, but otherwise equivalent to the  dense ones.
By construction, the spectrum of a single RMT block of linear size $N$ has support in $[-2\sqrt{N},2\sqrt{N}]$.
The supports of a collection of blocks therefore overlap in the $[-2\sqrt{N_\text{min}},2\sqrt{N_\text{min}}]$ region, where $N_\text{min} = \min_j N_j$ is the linear size of the smallest block in the collection.
In order to avoid boundary effects, we restrict the numerical computation of the level statistics to the central quarter of this overlap region. The normalized densities $\mu_i$ in that region are given by $\mu_i=\sqrt{\frac{N_i}{N_\text{tot}}}$ with $N_\text{tot} = \sum_j N_j$ the total linear size. 
In all numerical computations presented here, $N_\text{tot}$ is at least $2\times 10^3$, and $3.6\times10^5$ samples are used. 

Figure \ref{fig:rmt} displays the results of this comparison for $m=2,3,4$ (as well as the surmise of \cite{atas2013distribution} for $m=1$) and all Gaussian ensembles. The comparison is excellent and within the scale of this figure, there is no visible difference between the analytically obtained $P_m(r)$ and the numerical results $P_m^{\rm num}(r)$. More precisely, we found that the relative error $|P_m(r)-P_m^{\rm num}(r)|/P_m(r)$ is always less than $0.01$. This translates into also an almost perfect agreement (with no difference within error bars) between $\langle r \rangle_m$ (from Tab.~\ref{tabcoef}) and the numerical estimates reported in Tab.~\ref{tab:rgaps}. 

\begin{table}[ht]
\centering
\begin{tabular}{| c | c | c | c | c | c |}
\hline
  & $m$ & GOE      &   GUE & GSE  \\
  \hline
  $\langle r \rangle$ \cite{atas2013distribution} & 1 & 0.5307(1) & 0.5996(1) & 0.6744(1) \\
\hline
 \multirow{3}{*}{$\langle r\rangle_m$}& 2  & 0.4235(5) & 0.4220(5) & 0.4116(5) \\
 & 3 &  0.4035(5) & 0.3992(5) & 0.3927(5)\\
 & 4 &  0.3963(5) & 0.3924(5) & 0.3886(5)\\
 \hline
\end{tabular}
\caption{Values of averages $\langle{r}\rangle$ for $m$ blocks, as obtained from simulations on random matrices. The value for $m=1$ is taken from \cite{atas2013distribution}. Notice the excellent agreement with the surmise results reported in Tab.~\ref{tabcoef}.
\label{tab:rgaps}}
\end{table}

%%%%%%%%%%%%%%%%%%%%%
\begin{figure}[htp]
\includegraphics[width=.98\columnwidth]{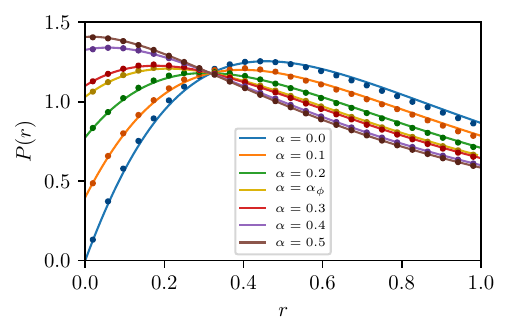}
\includegraphics[width=.98\columnwidth]{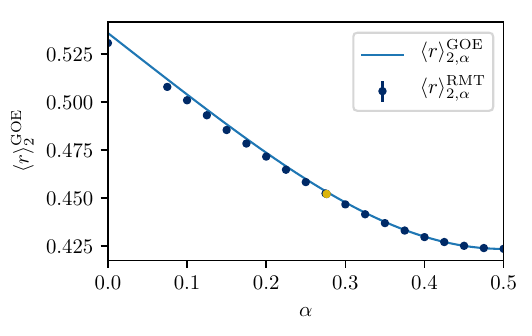}
\caption{Top: $P_2^\text{GOE}(r)$, for various density ratios $\alpha$ (see text). $\alpha_\phi = 1/(1+\phi^2)$ (in golden color in both plots) is the value that corresponds to the anyonic chain application discussed in Sec.\ \ref{sec:rsos}. Bottom: $\langle r \rangle_2^\text{GOE}$ as a function of $\alpha$. In both plots, full lines are predictions from the surmises of Sec.\ \ref{sec:analytics1}, and points are numerical results obtained on random matrices of size at least $N_\text{tot} = 2000$, averaged over $3.6\times10^5$ realizations.}
\label{fig:m2GOE}
\end{figure}
%%%%%%%%%%%%%%%%%%%%

We also compare analytical and numerical results for the specific case of $m=2$ GOE blocks of different sizes. In Fig.~\ref{fig:m2GOE} we present results for the probability distribution $P(r)$ for different values of $\alpha=\frac{\mu_1}{\mu_1+\mu_2}$ (top panel) as well as for the expectation value $\langle r \rangle_{2,\alpha}$ as a function of $\alpha$ (bottom panel). Here again the agreement between analytical and numerical results is striking. 
More precisely, we found that the relative error $|P_m(r)-P_m^{\rm num}(r)|/P_m(r)$ was always less than $0.01$ for block ratios $\alpha \geq 0.2$.
For $\alpha < 0.2$, the relative error increases, but remains below $0.05$.
The seemingly crossing point in Fig.~\ref{fig:m2GOE} is in fact not a crossing point, as one can convince oneself by using the exact expression for $P_2^\text{GOE}(r)$ and calculating its value with enough precision in the vicinity of that point.

We conclude this comparison section by discussing \cite{sun_color_2020}, which provides an analytical estimate for $P(r)$ for $m=2$, derived from a $4\times 4$ surmise which is forced to contain two levels of each of the $m=2$ blocks. This does not contain all possible patterns considered in Sec.~\ref{sec:jointPxy}. The estimated value for $\langle r \rangle$ obtained from this approach is approximately close to the one presented in Table~\ref{tabcoef} for the GOE, GUE but strongly differs for the GSE, while our results in this latter case agree with the numerical estimates in Table~\ref{tab:rgaps}.

\subsection{Symmetry detection in an experimental context}
\label{sec:smallsize}
The numerical comparisons in the previous subsection are done for matrix sizes $N \sim 2000$ and histograms are obtained from many random realizations. This allows to compare to the analytical results, obtained in the thermodynamic limit, with small enough statistical error bars. In the context of numerical simulations, the quantities $\langle r\rangle_m$ and $P_m(0)$ given in Table \ref{tabcoef} provide a signature allowing to identify the presence of a symmetry. This generalizes the existing results for $\langle r\rangle$ which is a quantity routinely used in numerical studies (see discussion in Sec.~\ref{sec:intro}) to identify the chaotic nature of a spectrum.

While most of the applications of gap ratio statistics have indeed been so far used inputs of numerical spectra of many-body systems, it is worth discussing applications to experimental spectroscopies, which typically involve less statistics (less realizations of disorder) and (sometimes) smaller spectra. For instance, the experimental measurement of $P(r)$ in Ref.~\cite{roushan_spectroscopic_2017} were performed in a system with $N=45$ energy levels, and using $4$ realizations of disorder. 
Other typical experiments probing disordered many-body quantum systems (often in the context of many-body localization) in various experimental platforms (cold atoms, trapped ions, superconducting qbits) average experimental results over $6$~\cite{schreiber_observation_2015,bordia_coupling_2016,bordia_periodically_2016,bordia_probing_2017,luschen_signatures_2017}, $12$~\cite{luschen_slow}, $20$~\cite{Guo_2020}, $24$~\cite{zhu2020probing}, $30$~\cite{smith_many-body_2016},  $50$~\cite{chiaro2020direct,Choi_2016}, and up to $197$~\cite{lukin_probing_2019,Rispoli_2019} realizations of disorder. Most of these platforms work on quantum systems with a minimum of tens of qbits or atoms with corresponding many-body spectra of at least $N=1000$ energy levels. In a different physical context, spectroscopy experiments on nuclei allow to resolve a quite large number of energy levels (often by combining results from different experimental techniques), typically from hundreds to thousands~\cite{nuclear_data_ensemble,Weidenmuller_RMP}.

Interestingly, already at sizes achievable experimentally our approach provides a signature of symmetries. As can be seen in Table \ref{tabcoef}, differences between values of $\langle r\rangle_m$ are quite small. Having experimental investigations in mind, we therefore propose to consider instead the quantity
\begin{equation}
\label{eqim}
I_m^{1/4}=\int_{0}^{1/4}P_m(r)dr,
\end{equation}
which is simply the integral of the distribution of $r$ up to a point chosen at $r=1/4$; this upper bound is arbitrary, but it is close to the crossing point $r\approx 0.288$ of $P_{m=2}^{\textrm{GOE}}(r)$ and $P_{m=3}^{\textrm{GOE}}(r)$. One can easily obtain a numerical estimate of $I_m^{1/4}$ from an experimental spectrum by counting the number of ratios less than $1/4$. From the analytical side, theoretical expressions can be obtained from our exact formulas and are given in Table \ref{tabim}.

\begin{figure}[!t]
\includegraphics[width=.98\columnwidth]{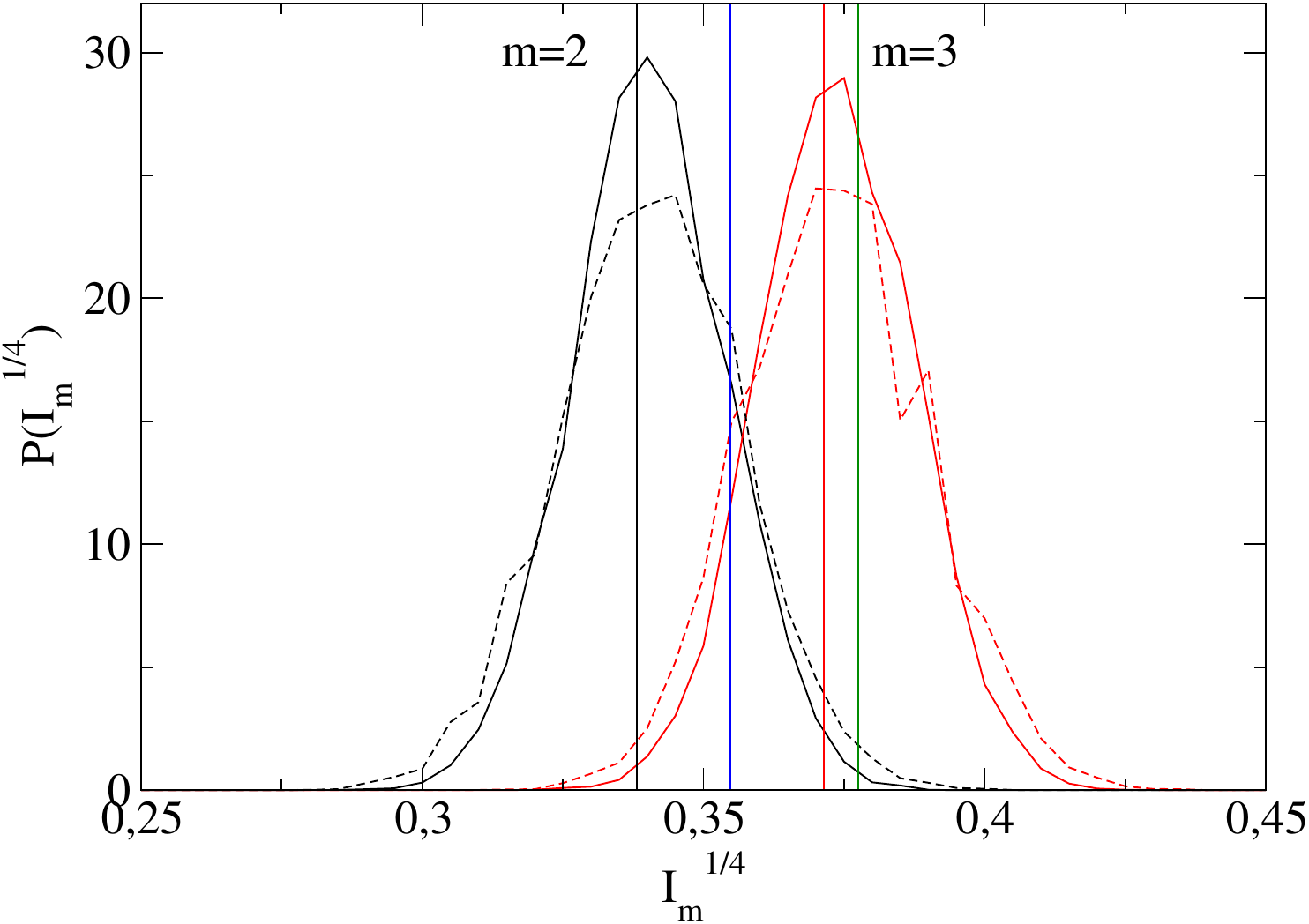}
    \caption{Probability distribution of $I_m^{1/4}$ for $N/m$ GOE blocks with $m=2$ (black) and $m=3$ (red). Histograms are obtained from 20000 values, each of which is calculated from 40 realizations of matrices of size $N=180$ (solid lines) and from 120 realizations  for size $N=48$ (dashed lines). Vertical black and red lines indicate the theoretical predictions for $m=2$ and $3$ respectively; blue line is the mid-value $\frac12(I_2^{1/4}+I_3^{1/4})$.}
\label{fig:smallsize}
\end{figure}

To illustrate this approach, we give an example of a `numerical experiment` where one would like to distinguish between the cases $m=2$ and $m=3$ in a case where the total number $N$ of available levels is small and realizations are scarce. In Fig.~\ref{fig:smallsize} we display probability distributions for the quantity $I_m^{1/4}$ when data are collected from spectra of size $N=180$ and when $40$ realizations of the experiment are available (solid lines in Fig.\ref{fig:smallsize}). In such a case, the number of available levels is very small since each block has size $N/m$ (90 for $m=2$ or $60$ for $m=3$). The two histograms associated with $m=2$ and $m=3$ are clearly distinguishable. In the case of an even smaller size $N=48$, one needs about $120$ realizations to get a comparable width of the histograms. These values of $N$ and number of realizations of disorder are comparable to the experimental situations discussed above.

More interesting is the probability of correct identification of the symmetry. If $m=2$, the experimentally measured value would be smaller than $\frac12 (I_2^{1/4}+I_3^{1/4})$ (the value which is equidistant from the $m=2$ and $m=3$ cases) in 89.2\% of cases. The criterion $I_m^{1/4}$ thus provides an additional tool, more suited to experimental situations where the number of realizations is scarce.

\begin{table}[ht]
\centering
\begin{tabular}{| c | c | c | c | c | c |}
\hline
  & $m$ & GOE      &   GUE & GSE  \\
  \hline
 \multirow{13}{*}{$I_m^{1/4}$}& 2  & 0.338171 & 0.250851 & 0.298583 \\
 & 3 &  0.37145 & 0.335806 & 0.363505\\
 & 4 &  0.383474 & 0.361911 & 0.377903\\
  & 5 &  0.389196 & 0.372592 & 0.382539 \\
  & 6 &  0.392374 & 0.3778 & 0.384388 \\
 & 7 &  0.394325 & 0.380654  & 0.385237 \\
   & 8 &  0.39561 & 0.382353 & 0.385667 \\
  & 9 &  0.396502 & 0.383429 & 0.385903 \\
  & 10 &  0.397146 & 0.384146 & 0.386039 \\
  & 11 &  0.397627 & 0.384641  & 0.386122 \\
  & 12 &  0.397996 & 0.384994 & 0.386175 \\
 & $\dots$ & $\dots$ &  $\dots$ &  $\dots$\\
 \cline{3-5}
 & $\infty$ (Poisson)  & \multicolumn{3}{c|}{0.4} \\
\hline
\end{tabular}
\label{tabim}
\caption{Value of $I_m^{1/4}$ defined by Eq.~\eqref{eqim} obtained from the surmise approach in Sec.~\ref{sec:analytics1}.}
\end{table}

\section{Illustrations in quantum many-body physics}
\label{sec:applications}

We now illustrate the usefulness of the above results by comparing them with simulations on realistic spectra obtained from quantum many-body problems. Most of our examples are taken from one-dimensional lattice models, mostly for computational convenience. In the following, the lattice will thus be a one-dimensional chain with $L$ sites.
Except otherwise mentioned, we will explicitly break translation symmetry, as well as possibly other lattice symmetries (such as reflection around the center of the chain) to concentrate on the existence of a few blocks.
The existence of translation symmetry would result in the existence of $L$ blocks (labeled by the $L$ reciprocal wave-numbers), which would result, as discussed earlier, in an (effective) Poisson distribution for level spacings and gap ratios in the thermodynamic limit.
The translation symmetry will be broken by using disorder characterized by a disorder strength $\epsilon$. In all the simulations presented below, we take $\epsilon$ not too small (in order to avoid the proximity to the translation-invariant case, which would cause stronger finite-size effects) as well as not too large, to avoid for instance a possible many-body localized phase (which would also result in Poisson spectral statistics). 
In all Hamiltonian systems we examine, we consider mid-spectrum eigenstates, obtained either by full diagonalization (for the smaller Hilbert space sizes) or by the shift-invert subset diagonalization method~\cite{pietracaprina_shift-invert_2018} for larger systems.

\subsection{Quantum clock models}

The first example deals with $Q$-states quantum clock models, which are natural $\mathbb{Z}_Q$-symmetric generalizations of the Ising quantum chain with $Q$-states quantum ``spins'' on each site \cite{Fendley_2012,Fendley_2014}. These exhibit a rich ground state phase diagram including ordered and disordered phases as well as critical lines, and have attracted a lot of attention in the recent years due to their relation with parafermions, a $\mathbb{Z}_Q$ generalization of Majorana fermions \cite{Fradkin_Kadanoff_1980}, as well as with topological phases \cite{Alicea_Fendley_2016}. They are furthermore related to cornerstone models of statistical mechanics, including the Potts model (where the $\mathbb{Z}_Q$ symmetry is promoted to a larger, $S_Q$ symmetry) and the chiral Potts model \cite{Albertini_McCoy_Perk_1989,Baxter_1988}.  

On each site, we define a spin taking $Q$ possibles values ($0 \dots Q-1$), as well as two operators $\sigma$ and $\tau$, which generalize the Pauli matrices $\sigma^z$ and $\sigma^x$ of the Ising chain: $\sigma$ measures the orientation of the spin, while $\tau$ rotates it by one unit ``around the clock'', and as a result these fulfill the following algebraic rules:  $\sigma^Q=\tau^Q=1$, $\sigma^\dag=\sigma^{Q-1}$, $\tau^\dag=\tau^{Q-1}$ and $\sigma \tau = \omega \tau \sigma$ with $\omega=\exp(2 i \pi/Q)$, a $Q$th root of unity.

Simple matrix representations are obtained in the basis where $\sigma$ is diagonal (the ``$\{\sigma\}$-basis''):
	\begin{align}
	\sigma= \left(
	\begin{array}{cccc}
	1 & & & \\
	  & \omega  & &\\
	    &  & \ddots &\\
	      &  & & \omega^{Q-1}
	\end{array}
	 \right)  \,,
	 \qquad
	 \tau = \left(
	\begin{array}{cccc}
	0 & 1 & &  \\
	  & \ddots  & \ddots  & \\
	    &  & \ddots &  1\\
	    1  &  & & 0
	\end{array}
	 \right) \, .
	 \label{tausigma}
	\end{align}
In the basis where $\tau$ instead is diagonal (the ``$\{\tau\}$-basis''), the matrices are exchanged.

The standard Hamiltonian for quantum clock models is written as a linear combination of $(\tau_j)^a$, $a=1,\ldots,Q-1$ on each site $j$ and exchange terms $(\sigma^\dag_j \sigma_{j+1})^a$, $a=1,\ldots,Q-1$. It is invariant under a $\mathbb{Z}_Q$ ``clock'' symmetry $\sigma_j \to \omega \sigma_j$, and the associated conserved charge $Z =\prod_j \tau_j$ has eigenvalues ${1,\omega,\ldots \omega^{Q-1}}$. For $Q\geq 3$ the original model has two other important symmetries: charge conjugation, which acts as $\tau_j \to \tau_j^\dagger$, $ \sigma_j  \to \sigma_j^\dagger$, and time-reversal, which is anti-unitary (and therefore sends any constant to its complex conjugate) and sends $\sigma_j$ to $\sigma_j^\dagger$ while leaving $\tau_j$ invariant. 

The model we consider in the following breaks all symmetries, but $\mathbb{Z}_Q$ :
\begin{equation}
	H_Q = - \sum_{j} J_j \sigma^\dag_j \sigma_{j+1} + \Gamma \tau_j + i g (\tau_j - \tau_j^\dag) \sigma_j \sigma_{j+1}^\dag + h.c.,
	\label{eq:potts}
\end{equation}
where the sum runs over the $L$ sites of the 1d lattice. 
For practical computations we restrict ourselves to $Q=2,3,4$. 
The coupling constants $J_j$ are independent random numbers uniformly taken from a box distribution $[J-\epsilon, J + \epsilon]$. Since they are a priori different on each site, they break invariance under translation or spatial reflection. 
The last term breaks both time-reversal and charge conjugation symmetry (this breaking could also have been achieved by perturbing with the $U(1)$ charge $S^z$ introduced in \cite{Vernier_U1_2019}). 

%%%%%%%
\begin{figure}[!h]
\includegraphics[width=0.98 \columnwidth]{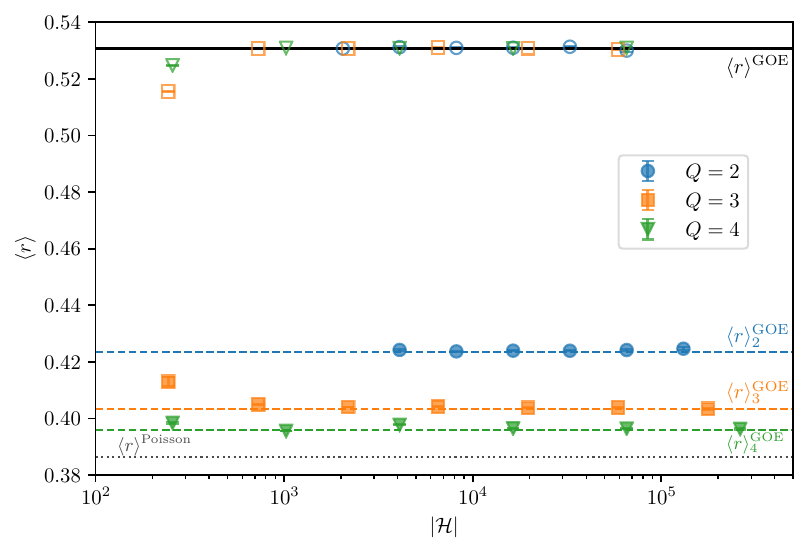}
\includegraphics[width=0.98 \columnwidth]{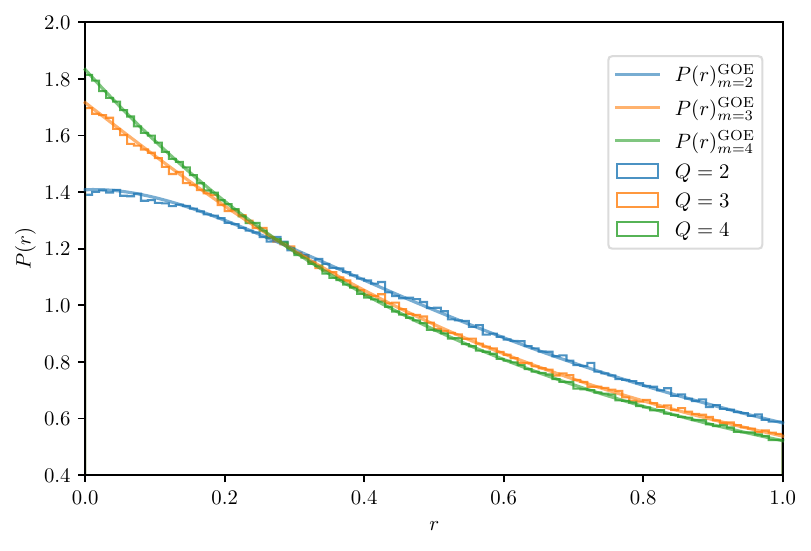}
\caption{{\it Q-states clock model --- } Top: Average gap ratio $\langle r \rangle$ for the model of Eq.~\eqref{eq:potts} for different values of $Q$, as a function of Hilbert space size $|\mathcal{H}|$. Open symbols denote simulations where the $\mathbb{Z}_Q$ symmetry is resolved, in which case the Hilbert space size is the block size $Q^{L-1}$ (we averaged data over all $Q$ equivalent blocks). Filled symbols denote results for the full Hilbert space of size $Q^{L}$. The dashed lines represent the values of $\langle r \rangle_{m}^{\mathrm {GOE}}$ obtained in Sec.~\ref{sec:analytics1} (taken from Table~\ref{tab:rgaps}), while  $\langle r \rangle^{\mathrm {GOE}}$ is the numerical estimate for the GOE distribution taken from \cite{atas2013distribution}. The precision of our numerics allows us to clearly distinguish the case $m=4$ blocks with $\langle r \rangle_{m=4}^{\mathrm{GOE}}$ from the Poisson value $\langle r \rangle^{\mathrm{Poisson}}$ also represented in the plot. Simulations parameters are $\epsilon=0.5$, $J=1$, $\Gamma=0.8$, $g=0.5$. For $Q=2$, instead of the time-reversal and charge conjugation breaking term in Eq.~\eqref{eq:potts} which vanishes when $Q=2$, we add a next-nearest neighbor interaction $g_2 \sum_j \sigma_j \sigma_{j+2}$ in order to break the mapping to a free-fermion model (we take $g_2=0.5$). Statistics are obtained by focusing on $200 Q$ eigenstates in the middle of the full spectrum, except for the smaller sizes where $\sim 20 Q$ eigenstates where considered. Results are averaged over more than $4000$ realizations of disorder, except for the largest size where $1000$ realizations were used. For $Q=2,3,4$, we obtained results on chains of sizes up to $L=17,11,9$ respectively. Bottom: Probability distribution of the gap ratio $P(r)$, as obtained from simulations of chains of sizes $L=16,10,8$ for $Q=2,3,4$ respectively. Simulation parameters are the same as in top panel. The solid lines represent the surmises $P_{m}^{\mathrm{GOE}}(r)$ obtained from the analytical computations in Sec.~\ref{sec:analytics1}.}
 \label{fig:potts}
\end{figure}
%%%%%%%

We first consider results of simulations performed in the $\{\sigma\}$ basis, with a full Hilbert space of size $Q^L$. In the top panel of Fig.~\ref{fig:potts} we present the average gap ratio for different chain sizes.
We clearly observe gap ratios which do not tend to their GOE value $\langle r \rangle^{\mathrm{GOE}}$, but rather to their $\langle r \rangle_{m=Q}^{\mathrm{GOE}}$ value as the size of the Hilbert space is increased. This is expected, as the Hamiltonian possesses $Q$ sectors of identical size $Q^{L-1}$ labeled with the different eigenvalues of the charge $Z$. Note however that working in the $\{\sigma\}$ basis does not allow to simply construct the Hamiltonian blocks, as $Z$ is off-diagonal in that case. Furthermore, the Hamiltonian is complex in this basis, and without any further indication on the existence of the $\mathbb{Z}_Q$ symmetry, it would not be clear why GUE statistics should not show up. 

When switching to the $\{\tau\}$ basis, $Z$ is now diagonal and the blocks are easily constructed. Furthermore the Hamiltonian becomes real. Computing the average gap ratio in each block leads to an asymptotic $\langle r \rangle^{\mathrm {GOE}}$ value for each block, showing that each block is indeed independent and no further symmetry has been missed.

We further confirm these results by showing the full distribution $P(r)$ in the bottom panel of Fig.~\ref{fig:potts} for $Q=2,3,4$. When the full spectrum is taken, the distribution obtained numerically for the largest system size is in excellent agreement with the surmises $P_{m=Q}^{\mathrm{GOE}}(r)$ obtained from Sec.~\ref{sec:analytics1}. 

Note the importance of the time-reversal symmetry breaking term $g\neq 0$ in Eq.~\eqref{eq:potts} in this analysis. In the presence of time-reversal symmetry (at $g=0$), the blocks with $Z$ and $Z^*$ are identical, leading to exact degeneracies. These additional values at $r=0$ would result in effective values of $\langle r \rangle$  lower than their Poisson values for finite-size systems.

\subsection{Discrete symmetries in disorder distributions}

In this section, we consider the Heisenberg spin chain in the presence of a random external field $h_j$:
\begin{equation}
    \label{eq:XXZ_model}
    H_\text{Heisenberg} = \frac{1}{2} \sum_{j=1}^L \bm{\sigma}_j \cdot \bm{\sigma}_{j+1} - \sum_{j=1}^L h_j \sigma_j^z,
\end{equation}
where $\sigma^{x,y,z}$ are the standard Pauli matrices, and we use periodic boundary conditions.
In general, this system hosts a many-body localized phase at large enough disorder: in particular, the model with box disorder has become the standard model of MBL in one dimension~\cite{pal_many_2010,luitz_many_2015}.
When disorder is reduced, the system undergoes a transition towards a thermal phase, a signature of which is an RMT-like spectral statistics.
Since uncorrelated disorder explicitly breaks all spatial symmetries, we expect the spectral statistics in the thermal phase to be of GOE type.

However, in the specific case of \emph{binary} disorder, i.e.\ taking on discrete values $h_j = \pm h$, visible deviations from the GOE gap ratio distribution were observed in the bulk of the thermal phase \cite{janarek_discrete_2018}.
The authors of \cite{janarek_discrete_2018} explained that this phenomenon was due to the peculiarities of discrete disorder distributions.
Indeed, while on a finite-size system a \emph{typical} disorder configuration will break all spatial symmetries, with discrete disorder distributions such as the binary one, there is a non-zero probability that one or several of them are preserved, specially when considering periodic boundary conditions.
For example, out of the $2^4 = 16$ possible binary disorder configurations on $L=4$ sites, 4 are reflection symmetric: $(+h,+h,+h,+h)$, $(+h,-h,-h,+h)$, \dots (actually, all disorder configurations on $L=4$ sites have a spatial symmetry: reflection, translation, inversion -- exchanging $h \leftrightarrow -h$, or a combination of them).
Of course, when $L$ is increased, the probability of drawing a spatially symmetric configuration decreases exponentially fast; but for the largest system sizes within reach of exact diagonalization techniques, the fraction of symmetric disorder configurations is still large enough to significantly alter the level statistics if disorder averaging is done ``naively", that is, by uniformly sampling over disorder.
A possible workaround put forward in \cite{janarek_discrete_2018} is to discard the spatially symmetric disorder configurations.
If one insists on using all samples, another possibility to is to explicitly resolve the symmetry block structure of the Hamiltonian, whenever the disorder configuration happens to be symmetric.
This is cumbersome, especially given the number of possible symmetries that must be taken into account.

%%%%%%%%%%%%%%%%%%%%%
\begin{figure}[h]
\includegraphics[width=.98\columnwidth]{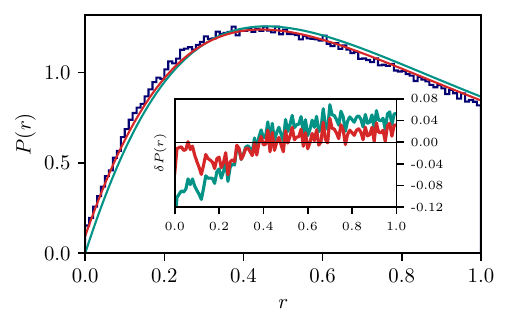}
\caption{Probability distribution of the gap ratio from the Heisenberg model of Eq.~\eqref{eq:XXZ_model} with $L=18$ spins, with periodic boundary conditions. Data is averaged over all realizations of the binary transverse field $h_j = \pm 1/2$ ($2^{18}$ in total, 3914 of which are nonequivalent up to symmetries), and over 150 eigenstates around infinite temperature energy $E = (E_\text{min} + E_\text{max})/2$. The red solid line represents the analytical prediction Eq.~\eqref{eq:Pr_XXZ}: a linear combination of surmises $P_{m}^{\mathrm{GOE}}(r)$ obtained from the analytical computations in Sec.~\ref{sec:analytics1}. For comparison, the green solid line shows the predicted distribution when two blocks contributions are not taken into account. Inset: Difference $\delta P(r)$ between the numerical data and these surmises.}
\label{fig:XXZ}
\end{figure}
%%%%%%%%%%%%%%%%%%%%

In that context, using the surmise for several GOE blocks proves useful.
Indeed, the gap ratio in the thermal phase of the model can be written as a sum over symmetry sectors:
\begin{equation}
    \label{eq:Pr_XXZ}
    P(r) = \sum_{m} w_m P_m^\text{GOE}(r),
\end{equation}
where $w_m$ is the weight of the symmetry sector of $m$ blocks.
In the thermodynamic limit $L\to \infty$, $w_1 \to 1$, while $w_{m>1}$ decays exponentially fast to zero.
Note that for samples with two blocks ($m=2$) the two blocks are always of the same size, whereas for samples with more than 2 blocks, $m > 2$, blocks are not necessarily of equal size. While the expressions in the previous section allow us to compute the surmise for these non-homogeneous samples, we can to a good degree of approximation neglect their contribution to the gap ratio distribution. Indeed, we find using simple combinatorics, that the total weight $\sum_{m>2} w_m$ coming from samples with more than two blocks is more than halved when $L \to L+2$, and for $L=18$, it already represents less than $0.2\%$ of the total weight.
We will therefore make the approximation that $P(r) = w_1 P_1^\text{GOE}(r) + w_2 P_2^\text{GOE}(r)$.

In Fig.\ \ref{fig:XXZ}, we show the gap ratio distribution for the Hamiltonian Eq.~\eqref{eq:XXZ_model} for  $L=18$. We find $w_1 = 243936/2^{18} \simeq 0.93$, $w_2 = 17640/2^{18} \simeq 0.07$.
This system size is of the order of what is achievable using state-of-the-art exact diagonalization techniques targeting the middle of the energy spectrum \cite{pietracaprina_shift-invert_2018}.
However, it is not large enough for $w_2$ to be negligible compared to $w_1$.
Indeed, as shown in Fig.\ \ref{fig:XXZ}, incorporating the $m=2$ contribution visibly improves the agreement with numerical data. Note the clear difference at $r=0$ between Eq.~\eqref{eq:Pr_XXZ} (for which $P(0)\neq 0$, as in the numerical simulations of Eq.~\eqref{eq:XXZ_model}) and $P^{\mathrm{GOE}}(r)$ which vanishes at $r=0$ due to level repulsion. Accordingly, the predicted average gap ratio using the $m=2$ surmise $\langle r \rangle = w_1 \langle r \rangle_1 + w_2 \langle r \rangle_2 \simeq 0.527$ is closer to the numerically computed value $\langle r \rangle - \langle r \rangle_\text{Heisenberg} \simeq 0.004$ than the ``naive" prediction involving only $m=1$: $\langle r \rangle_{m=1} - \langle r \rangle_\text{Heisenberg} \simeq 0.013$.

\subsection{Floquet spin chain model}

We next consider both a static and a Floquet spin $1/2$-chain model.
Floquet systems have attracted a great deal of interest, because while they are amongst the simplest non-equilibrium Hamiltonian systems, they exhibit new non-trivial properties, that are not observed in their static cousins. In particular, single-particle Floquet systems can host topological phases that have no static equivalent \cite{review_noninteracting_floquet}.
Interacting many-body Floquet systems a priori exhibit no such interesting phases of matter, since the combination of interaction and driving is expected to heat up the system to an infinite-temperature, featureless state \cite{dalessio_long_2014, lazarides_coe_2014, ponte_coe_2015}.
However, it has been shown \cite{ponte_mbl_floquet, lazarides_mbl_floquet, abanin_mbl_floquet, harper_review_interacting_floquet} that the addition of disorder, hindering energy propagation throughout the system via a MBL mechanism, can prevent heating and give rise to new interacting Floquet phases, such as discrete time crystals \cite{yao_dtc, khemani_phase_2016}. Here, we study an interacting Floquet system, along with its static counterpart, for comparison.
We show in the following that the Floquet system exhibits an extra symmetry, that can be associated to Floquet topological modes.
In order to detect the symmetry, we adjust the system parameters so as to be in the thermal phase of the model. Then, level statistics is expected to follow RMT predictions, enabling us to employ our surmises to detect the Floquet symmetry.

We work with the following spin $1/2$ Hamiltonians:
\begin{align}
    H_x &= \sum_{j=1}^L g \sigma_j^x \nonumber \\
    H_z &= \sum_{j=1}^L J \sigma_j^z \sigma_{j+1}^z + \sum_{k=0}^{L/2-1} h_{2k+1} \sigma_{2k+1}^z,
\label{eq:spin}
\end{align}
again with periodic boundary conditions, 
which we combine to form a time-independent $H^\text{static} = H_x + H_z$ and a time-dependent model:
\begin{equation}
    H^\text{driven}(t) = \begin{cases} 2H_z &\mbox{if } 0 \leq t\!\!\!\mod \tau < \tau/2 \\
2 H_x & \mbox{if } \tau/2 \leq t\!\!\!\mod \tau < \tau \end{cases}.
\end{equation}
Because the drive is periodic $H^\text{driven}(t + \tau) = H^\text{driven}(t)$, such a model is indeed a Floquet system.

In the Floquet setting, energy is not conserved. It is replaced by \emph{quasi-energy}, which is defined up to arbitrary shifts by $2 \pi / \tau$.
More specifically, let us introduce the Floquet operator $U_F = \exp(-i \tau H_x) \exp(-i \tau H_z)$, which is the evolution operator over one drive period.
To the unitary Floquet operator we can associate a Floquet Hamiltonian $H_F$, defined as $U_F = \exp(-i \tau H_F)$, whose eigenvalues $\varepsilon_\alpha$ and associated eigenvectors are respectively the quasi-energies and the Floquet eigenstates, which hold information about the dynamics and steady-state properties of the system \cite{shirley_floquet_theory}.
In practice, when computing level statistics, we will therefore use the quasi-energies $\varepsilon_\alpha$ exactly like energies in the static case.

Going back to the system Eq.~\eqref{eq:spin}, remark that the random longitudinal fields $h_j \neq 0$ break both the Ising and the translation symmetries.
Our model differs from the most commonly used one in that $h_j = 0$ on even sites. This does not change the physics of the model, but can induce an extra symmetry in the driven case, as we discuss below.
Finally note that driving the system does not break its time-reversal symmetry.
We therefore expect a GOE (respectively COE) level statistics in the time-independent (respectively driven) case \cite{dalessio_coe_2014, lazarides_coe_2014, ponte_coe_2015, kim_eth_2014}. Since the COE and GOE ensembles are asymptotically described by the same statistics, we will compare simulations in the driven case to the corresponding GOE statistics.

\begin{figure}[htp]
\includegraphics[width=0.98\columnwidth]{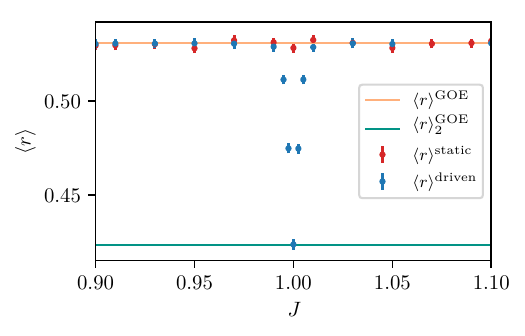}
\includegraphics[width=0.98\columnwidth]{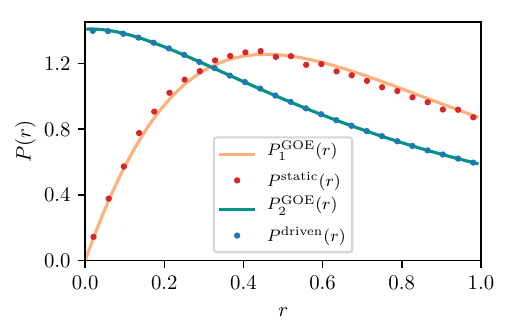}
\caption{Top: Average gap ratio as a function of $J$ in the time-independent and driven spin-$1/2$ chain model Eq.~\eqref{eq:spin}. Bottom: Gap ratio distributions at $J=1$ for the time-independent and driven case (points), and comparison with the surmise distribution for $m=1$ and $m=2$ GOE blocks (full lines). In the time-independent case, average is performed over 2000 disorder realizations (except at the $J=1$ point where 5000 realizations are used), and 100 eigenstates around infinite temperature energy $E = (E_\text{min} + E_\text{max})/2$, for the system of size $L=14$, while in the driven case, average is performed over 2000 disorder realizations and all eigenstates of the system of size $L=12$.}
\label{fig:floquet}
\end{figure}

We choose the parameter set $g = \Gamma \times 0.9045$, $h_{2 k + 1} = 0.809 + 0.9045 \times \sqrt{1 - \Gamma^2} \epsilon_k $, $\Gamma = 0.9$, $\tau = \pi/4$, where the $\epsilon_k$ are uniformly distributed random number of zero mean and unit variance. This choice of parameters has been shown \cite{zhang_thermalization_2015, zhang_mbl_2016, lezama_slow_2019} to give good agreement with COE level statistics on the accessible system sizes, for the related model where the longitudinal field is also non-zero on the even sites: $h_{2k} \neq 0$.
This is indeed the case for the time-independent system, as can be seen in Fig.~\ref{fig:floquet}. However, the top panel of Fig.~\ref{fig:floquet} shows that there is a dip in the average gap ratio $\langle r \rangle$ around $J=1$ for the driven system.
The numerical estimate for $\langle r \rangle$ at this special point appears to coincide with the surmise value for two GOE blocks of equal size given in Table \ref{tabcoef}.
The level statistics (bottom panel of Fig.~\ref{fig:floquet}) is also compatible with the gap ratio statistics $P_{m=2}^{\mathrm{GOE}}(r)$. Driving the otherwise fully GOE system therefore appears to give rise to a new $\mathbb{Z}_2$ dynamical symmetry at the $J=1$ point.

%%%%%%%%%%%%%%%%%%%%%
%%%%%%%%%%%%%%%%%%%%
 
We now rationalize the emergence of this symmetry. We find that the associated conserved operator is
\begin{equation}
    X = \prod_{k=0}^{L/2 - 1} \sigma_{2k+1}^x.
\end{equation}
Indeed, while this operator acts in general non-trivially on the $H_z$ Hamiltonian, we have at the special point $J=1$,
\begin{equation}
\label{eq:floquet_commutator}
    X e^{i H_z} X e^{-i H_z} = \prod_{j=1}^Li \sigma_j^z\sigma_{j+1}^z = (-1)^{L/2},
\end{equation}
and thus $X$ commutes with the Floquet operator, up to a global phase factor which can be absorbed in the definition of $U_F$. This indicates the existence of 2 COE blocks of identical sizes in the Floquet Hamiltonian, and hence explains the agreement with the analytically-obtained gap probability distribution $P_2^{\mathrm{GOE}}(r)$.
Note that setting $h_j = 0$ on odd (or even) sites is necessary for this extra symmetry to exist.
Remark that when open boundary conditions are used, the above commutator Eq.~\eqref{eq:floquet_commutator} becomes proportional to $\sigma_1^z \sigma_{L}^z$, a non-trivial boundary term.
We can interpret this boundary term as creating a pair of excitations \cite{berdanier_criticality}.
If the model were brought to the MBL regime (e.g.\ by increasing the strength of the disorder term $h_j$), these excitations would become localized at both ends of the chain, signalling the topological nature of the observed $\mathbb{Z}_2$ symmetry.
However, in that case the level statistics would become Poisson, and we would not be able to detect the symmetry using our RMT approach.
Finally, we note that the argument carries over when we add a third contribution $H_y = \sum_{k=0}^{L/2-1} h_y \sigma_{2k+1}^y$.
It breaks the time-reversal symmetry, turning the 2-block COE structure of the $J=1$ point into a 2-block CUE structure.

\subsection{Anyonic Chain}
\label{sec:rsos}

\begin{figure}[h]
\includegraphics[width=0.98 \columnwidth]{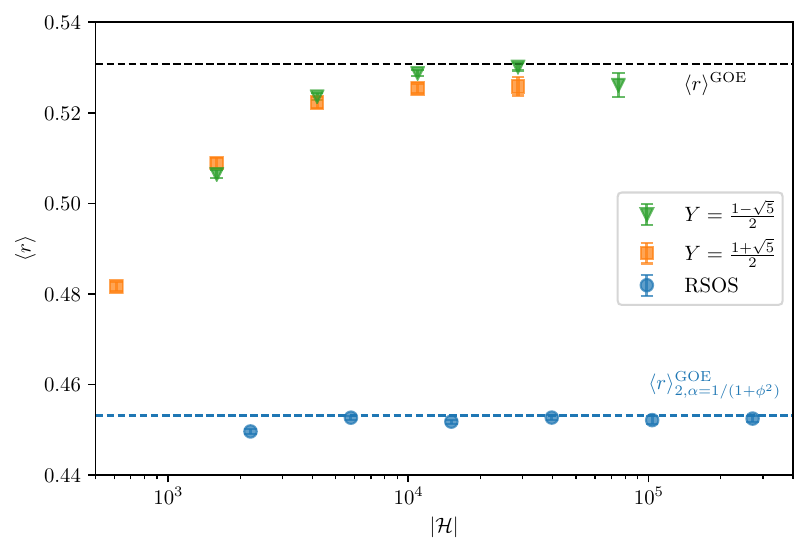}
\includegraphics[width=0.98 \columnwidth]{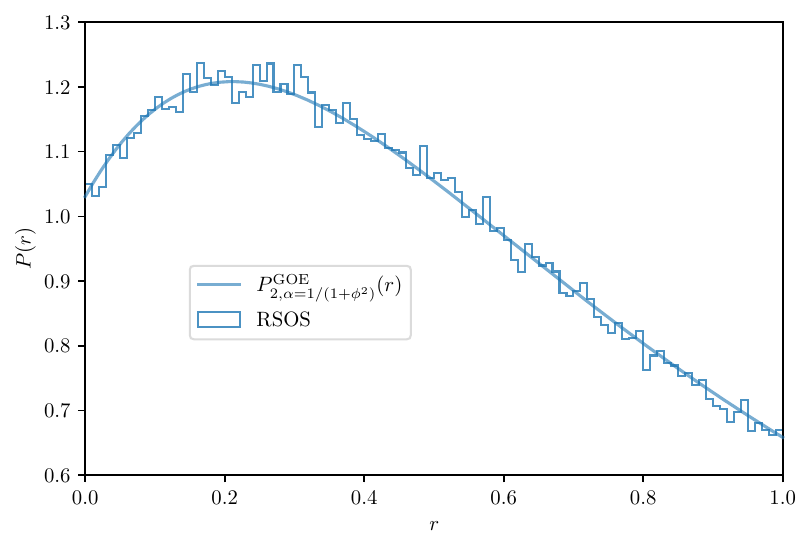}
\caption{{\it RSOS model --- } Top: Average gap ratio $\langle r \rangle$ for the RSOS model Eq.~\eqref{HRSOS}, as a function of Hilbert space size. Blue circles are the results for the full spectrum of the Hamiltonian \eqref{Hanyons}; squares and triangles are those in the $Y=\frac{1\pm \sqrt{5}}{2}$ sectors. The dashed lines represent the results for the (single-block) GOE distribution of~\cite{atas2013distribution} and from the surmise computations in Sec.~\ref{sec:analytics1} for $m=2$ GOE blocks with size ratio $\phi^{-2}$. Data in the $Y=\frac{1+ \sqrt{5}}{2}$ sector were obtained by comparing energy spectrum in the loop representation (with non-contractible loop weight $2 \cos{\pi/5}$) and RSOS representation. Data in the $Y=\frac{1- \sqrt{5}}{2}$ were obtained by considering the rest of the states in the RSOS representation. We focus on mid-spectrum eigenstates of the RSOS spectrum, obtaining $\sim 300$ eigenstates for every disorder realization. Results are averaged over between $300$ and $1000$ realizations of disorder of strength $\epsilon=0.2$. Bottom: Probability distribution of the gap ratio $P(r)$, as obtained from simulations of a RSOS chain of size $L=22$. The solid line is the surmise $P_{m=2,\alpha=1/(1+\phi^{2})}^{\mathrm{GOE}}(r)$ obtained from the analytical computations in Sec.~\ref{sec:analytics1}.}
 \label{fig:rsos}
\end{figure}

Our final application deals with chains of interacting anyons, which are exotic particles interpolating between bosonic and fermionic statistics. They are predicted to occur in some two-dimensional systems such as fractional quantum Hall states \cite{ReadRez,Bartolomei2020}, and offer exciting perspectives for topological quantum computation \cite{Nayak}. 
More precisely, we will consider a disordered version of the ``golden chain'' model of Fibonacci anyons~\cite{goldenchain}. 
As more technical background is needed to introduce the model and its various representations, we first give a summary of our results. When periodic boundary conditions are imposed on the chain of anyons, there is a non-trivial topological symmetry that decomposes the Hamiltonian into two blocks of unequal size. Resolving this symmetry is not easy, but it can in principle be done at the price of turning the Hamiltonian into a dense matrix, rendering numerical simulations on large systems difficult. Our results instead allow to use a representation simpler for numerics (with sparse, real symmetric matrices) which can nevertheless be confronted to RMT predictions, and hence probe ergodic physics. This can be seen in Fig.~\ref{fig:rsos} where the results for $\langle r \rangle$ and $P(r)$ allow to characterize the spectral statistics of the model with the two interlaced sectors. At an extra numerical cost, and with the further requirement to study different representations of the model, we can identify the states in each of these two sectors and check that they follow regular single-block GOE statistics (squares and triangles in the top panel of Fig.~\ref{fig:rsos}). In the following, we present in detail the different representations of the model of disordered Fibonacci anyons, which allow us to draw these conclusions.

The statistics of anyons \footnote{A nice introduction can be found in the online caltech course of J.  Preskill, Chapter 9 - Topological Quantum Computation, available at \url{http://www.theory.caltech.edu/~preskill/ph219/}} are generally encoded in a set of fusion rules analogous to the composition rules for angular momenta, as well as transformation rules relating the different possible ways to fuse together three or more particles (the so-called ``F-symbols'') \cite{Bonderson_2007}. In the case of Fibonacci anyons there are only two types of particles, the trivial particle, labeled by $1$, and the Fibonacci anyon, labeled by $\tau$.
They are characterized by the fusion rule $\tau \times \tau = 1 + \tau$, that is,  bringing together two Fibonacci anyons yields either the trivial particle or another Fibonacci anyon. This is analogous to the situation where two spin-$\frac12$ particles brought together can be decomposed into a spin-0 and a spin-1 particle.
In addition the trivial fusion rules $1\times 1 =1$, and $1 \times \tau = \tau$ hold \cite{goldenchain}.

Suppose now we have a chain of $L$ indistinguishable Fibonacci anyons. A pair of adjacent anyons may be fused together, yielding either 1 or $\tau$. Performing recursively all possible fusions, we end up with a single anyon, again either 1 or $\tau$. The different ways by which the $L$ particles pair up and fuse to yield a single particle has the structure of a Hilbert space and is called the fusion space.
In contrast with the case of spins, this Hilbert space does not have a tensor product structure. In order to construct a basis for this Hilbert space it is convenient to consider the different ``fusion paths'' which describe the outcome of each fusion, starting from the leftmost pair (particle 1 with particle 2, then the resulting particle with particle 3, and so on). Each fusion path can be written as a sequence $| x_1 x_2 \ldots x_L \rangle$, where for each $i$, $x_i \in \{1,\tau\}$, and $x_{i+1}$ is the outcome of the fusion of $x_i$ with $\tau$ ($x_1$ being the outcome of the fusion of the first two particles). Since the fusion of 1 with $\tau$ always yields $\tau$, no two consecutive $1$s are allowed in the sequence of $x_i$. In fact, the basis is given by all strings which do not contain any pair of consecutive $1$s.

In the case of periodic boundary conditions, $x_{L+1}=x_1$, and the number of basis states $|\cal{H}|$ is related to the Fibonacci sequence, $|{\cal H}(L)|= F_{L-1}+F_{L+1}$, where $F_L$ is the $L$th Fibonacci number with $F_0=0$ and $F_1=1$. It is well known that the ratio of consecutive Fibonacci numbers goes to $\lim_{i \rightarrow \infty} F_{i+1}/F_i = \phi$  with $\phi=\frac{1+\sqrt{5}}{2}$ the golden mean, hence the name golden chain. A pedagogical introduction to the Hilbert space and Hamiltonian construction of the golden chain model can be found in \cite{trebst_anyons}.

Following the seminal work \cite{goldenchain}, a Hamiltonian can then be constructed by assigning a different energy for each possible outcome of the fusion of two nearest neighbors at sites $j$ and $j+1$. Assigning a zero energy to an outcome 1 and $-J_j$ to an outcome $\tau$,  the Hamiltonian takes the form
\begin{equation}
H = - \sum_j J_j \Pi_{j,j+1} \,,
\label{Hanyons}
\end{equation}
where $\Pi_{j,j+1}$ is the projector into the trivial particle of two $\tau$ particles located at sites $j$ and $j+1$. This is the analog of the Heisenberg coupling for $SU(2)$ spins 1/2, which assigns a different energy to the fusion channels of pairs of neighboring spins. The projector $\Pi_{j,j+1}$ acts on a basis state $| x_1 x_2 \ldots x_L \rangle$ by changing $x_j$ to a superposition of $1$ and $\tau$ in a way depending on $x_{j-1}$ and $x_{j+1}$; an explicit expression can be found in \cite{goldenchain}. The coupling constants $J_j$ are taken from a random distribution $P(J)=\epsilon^{-1} J^{-1+1/\epsilon} \theta(J)\theta(1-J)$ with $\theta$ the Heaviside step function ($\epsilon \in [0,\infty]$ characterizes the disorder strength and $J\in [0,1]$). Once again, the main interest for using a disordered coupling constant is to break lattice symmetries in the chain. We use periodic boundary conditions in the following.

A practical representation for numerical simulations is to recast the above fusion paths in terms of sequences of heights $| h_1 h_2 \ldots h_L \rangle$, where $h_i \in \{1,2,3,4\}$ and  $|h_i - h_{i+1}| = 1$, through the mapping $1,\tau \to 1,3$ for $i$ odd, $1,\tau \to 4,2$ for $i$ even. This defines a ``restricted solid-on-solid'' (RSOS) model, namely the $p=4$ case of the $A_p$  (also known as SU$(2)_{p-1}$) family, where for generic $p$ the heights are allowed to run between $1$ and $p$ \cite{Pasquier,Andrews_Baxter_Forrester_1984}.  In this formulation the  projectors $\Pi_{j,j+1}$ of Eq.~\eqref{Hanyons}  can be re-expressed in terms of operators $e_j$, whose action is defined as 
 \begin{align}
	& e_j | h_1 \ldots h_{j-1} h_j h_{j+1} \ldots h_L \rangle =
	 \nonumber\\
	 & \delta_{h_{j-1},h_{j+1}} \sum_{h_{j}'} 
	 \frac{ \sqrt{\sin(\frac{\pi h_j}{p+1} ) \sin(\frac{\pi h_j'}{p+1})}}{\sin (\frac{\pi h_{j+1} }{p+1} )}  | \ldots h_{j-1} h_j^{'} h_{j+1} \ldots  \rangle \,.
 \end{align}
The operators $e_j$ form a representation of the Temperley-Lieb (TL) algebra \cite{Temperley_Lieb_1971}, namely $e_j^2 = \sqrt{Q} e_j$, $e_j e_{j\pm 1} e_j = e_j$, and $e_i e_j = e_j e_i$ for $|i-j| \geq 2$, where we have defined $\sqrt{Q} = 2 \cos \frac{\pi}{p+1}$.  In the present case $p=4$, and one indeed checks that $\Pi_{j,j+1} = \frac{1}{\sqrt{Q}} e_j$. 
Up to the irrelevant $1/\sqrt{Q}$ proportionality factor, the Hamiltonian Eq.~\eqref{Hanyons} is therefore re-expressed in the RSOS representation as 
\begin{equation}
H_{\rm RSOS}= - \sum_j J_j e_j \,.
\label{HRSOS}
\end{equation}
A subtlety to keep in mind is that the RSOS formulation acts separately on two equivalent sectors, which correspond to putting even or odd heights on even sites respectively. For periodic boundary conditions, $h_{L+1}=h_1$, each of these sectors has size $F_{L+1}+F_{L-1}$, and yields a copy of the original anyonic chain. The spectrum of the original Hamiltonian Eq.~\eqref{Hanyons} is therefore obtained by restricting to a single of these sectors (which is what we do in the following).

The Hamiltonian Eq.~\eqref{HRSOS} is real, a reason for which this representation is often used in numerics. We present in Fig.~\ref{fig:rsos} (blue circles) the results for the average gap ratio $\langle r \rangle$ and its distribution $P(r)$ for Eq.~\eqref{HRSOS}, for different chain sizes (and thus Hilbert space sizes $|{\cal H(L)}|$) and weak disorder $\epsilon=0.2$, for states located in the middle of the spectrum and corresponding to the sector with even heights on even sites. For this small value of disorder, we do expect a random matrix theory behavior, but the value of $\langle r \rangle$ is clearly different from the GOE statistics for a single block. The size of the Hilbert space, which is the sum of two Fibonacci numbers, may suggest the existence of two blocks of different sizes (denoted $N_1$ and $N_2$ in the following). A first simple test is to compare the expectation value $\langle r \rangle_{\mathrm{RSOS}} \simeq 0.452$ to the values obtained for two GOE blocks of different sizes (Fig.~\ref{fig:m2GOE} in Sec.~\ref{sec:numerics}). This leads to a possible value around $\alpha=\frac{N_1}{N_1+N_2} \in [0.27,0.3]$, corresponding to a size ratio $N_1/N_2 \in [0.37,0.43]$, close to $\phi^{-2}=\lim_{L \rightarrow \infty} \frac{F_{L-1}}{F_{L+1}}\simeq 0.382$. This strongly suggests that the spectrum of the periodic RSOS chain is composed of two independent GOE blocks of size $N_1=F_{L-1}$ and $N_2=F_{L+1}$. 
In the top panel of Fig.~\ref{fig:rsos}, we also represent the predicted value $\langle r \rangle_{m=2,\alpha=1/(1+\phi^{2})}=0.453186$, to which the numerical data indeed appear to tend.  This is further confirmed by the numerical distribution of $P(r)$ (bottom panel of Fig.~\ref{fig:rsos}) which is an excellent match with the one obtained from the surmise (see Sec.~\ref{sec:analytics1}) of two GOE blocks with ratio $\phi^{-2}$.

%%%%%%%
%%%%%%%

We can in fact trace back this decomposition to the existence of a ``hidden'' symmetry of a topological nature \cite{goldenchain,anyonsspin1,Buican_Gromov_2017,topRSOS_2020}, namely an operator $Y$ corresponding to an extra $\tau$ particle circling around the system, and whose matrix elements in the basis of fusion paths may be written as 
\begin{equation}
\langle x'_1 \ldots x'_L  | Y | x_1 \ldots x_L  \rangle 
= \prod_{i=1}^{L} \left( F_{\tau x_i \tau}^{x_{i+1}'}\right)_{x_{i-1}}^{x'^{i}} \,, 
\end{equation}
where the F-symbols $\left( F_{\tau x_i \tau}^{x_{i+1}'}\right)_{x_{i+1}}^{x'_{i}}$ can be found for instance in \cite{goldenchain}.
The operator $Y$ has two distinct eigenvalues $\frac{1}{2}(1 \pm \sqrt{5})$, and commutes with the Hamiltonian Eq.~\eqref{Hanyons}, therefore defining two symmetry subspaces. A subtlety arises in the RSOS representation, which as discussed above has two independent sectors and where the action of $Y$ maps one onto the other. We can overcome this difficulty by computing the action of $Y^2$, which acts separately in the odd and even sectors: this allows to define in each sector two orthogonal subspaces of dimensions $F_{L+1}$ and $F_{L-1}$ respectively, which precisely reproduces the numerical observations made above \footnote{On a more algebraic note, the decomposition of the RSOS space into $F_{L+1}$ and $F_{L-1}$ can be understood as a decomposition into irreducible representations of the periodic Temperley-Lieb algebra \cite{pasquier_common_1990}. More recently, it has been observed that the two orthogonal subspaces of the periodic RSOS model can be obtained from the {\it open} chain with fixed boundary conditions, through an operation called {\it braid translation} \cite{braidtranslation}.}. 

Now, it is important to remark that the action of $Y$ is highly non-local, and its matrix expression in the RSOS representation is not sparse. Therefore while we know in principle how to decompose the Hamiltonian into two GOE blocks, it is not possible to our best knowledge to do so while keeping it sparse and real.  
One may ask whether other representations of our model might help with this problem. There are indeed other ways to represent the TL algebra, from which the spectrum of Eq.~\eqref{Hanyons} can be recovered. Below we consider two such representations, the loop representation and the spin chain representation. These representations allow us to tell apart which subspace each eigenvalue corresponds to.

\paragraph*{Loop representation}
In the loop representation \cite{Temperley_Lieb_1971}, the Hilbert space is spanned by the configurations of non-crossing valence bonds between $L$ vertical strands, and the TL generator $e_i$ acts by contracting together the strands at site $i$ and $i+1$. The composition rules of the TL algebra express the fact that lines can be continuously deformed without crossing, and that closed loops contribute a weight $\sqrt{Q}$. From there, one can recover the eigenvalues of the anyon chain corresponding to each symmetry sector by assigning a special weight to {\it non-contractible} loops which close around the cylinder \cite{SaleurBauer}, respectively $2\cos \frac{\pi}{5}$ and $2\cos \frac{2\pi}{5}$, which is nothing but the corresponding eigenvalue of $Y$. However, the loop model contains significantly more states than the anyonic chain, as the loops carry additional non-local information which is absent in the RSOS representation. This brings several difficulties, the first being that the maximum size $L$ that can be reached using exact diagonalization techniques is lower, the second being that it is not obvious at all how to extract from the loop model spectrum the set of eigenvalues which are present in the RSOS one \cite{Vernieranyons}. Furthermore, the  loop representation leads to a non-Hermitian matrix representation of the Hamiltonian, which also decreases the efficiency of simulations.

\paragraph*{Spin chain representation}
Another representation is in terms of a spin $1/2$ chain, with Hilbert space $(\mathbb{C}^2)^{\otimes L}$, on which the TL generators act as \cite{pasquier_common_1990} 
\begin{align}
    e_i = - &\left(e^{i  \frac{\varphi}{L}} \sigma_i^+ \sigma_{i+1}^- + e^{-i \frac{\varphi}{L}} \sigma_i^- \sigma_{i+1}^+ 
    + \frac{\cos \gamma}{2} (\sigma_i^z \sigma_{i+1}^z-1)  
    \right.
    \nonumber \\ 
    &
    \left. - \frac{i \sin \gamma}{2}  (\sigma_i^z - \sigma_{i+1}^z) 
    \right) \,.
\label{TLspin}
\end{align}
Here the matrices $\sigma^{x,y,z}_i$ act as Pauli matrices on the $i$th spin, and as identity elsewhere, and $\gamma$ is defined by $\sqrt{Q} = 2 \cos \gamma$. 
The role of the twist parameter $\varphi$ is analog to that of the weights of non-contractible loops in the geometrical representation. More precisely, the Hamiltonian built out of Eq.~\eqref{TLspin} commutes with the global magnetization $S^z=\sum_i \sigma_{i}^z$, and the eigenvalues of the RSOS model are recovered in the $S^{z}=0$ sector upon setting $\varphi = \frac{\pi}{5}$ and $\varphi=\frac{2\pi}{5}$, respectively. As for the loop case, the $S^z=0$ sector however contains more states than the RSOS ones, leading to the difficulties mentioned above (see \cite{Saleur,JacobsenSaleur,VJSalas} for other occurrences in related models). Moreover this Hamiltonian is complex in the $\sigma^z$ basis, which also leads to a decreased numerical efficiency. 

We use simulations both in the loop and spin chain representations and checked on small systems (up to $L=18$) that all states in the RSOS representation can indeed be found in the loop representation (using non-contractible loop weight taking either $2\cos \frac{\pi}{5}$ or $2\cos \frac{2\pi}{5}$ values) or the spin chain representation (using a twist taking either $\pi/5$ or $2\pi/5$ values). The simulations in the loop model with non-contractible loop weight $2\cos \frac{\pi}{5}$ are simpler (as all loops have the same weight) and we could reach larger systems. This allowed us to identify all states in the $Y=\frac{1+ \sqrt{5}}{2}$ sector for chains of size up to $L=24$ (see Fig.~\ref{fig:rsos}).

Besides allowing to identify this two-block structure in $H_{\rm RSOS}$ chains with periodic boundary conditions, the actual value of $\langle r \rangle$ and distribution $P(r)$ for $N_1/N_2=\phi^{-2}$ will be useful as a marker of an ETH/ergodic phase when increasing the value of disorder. Indeed, it has been proposed~\cite{vasseur_quantum_2015} that disordered SU(2)$_3$ chains could lead to a new form of non-ergodic, critical, phase which behavior is different from a many-body localized phase. This putative new critical phase could be identified by the departure of spectral statistics from the references values displayed above.

\section{Summary, relation to previous works and perspectives}
\label{sec:discussion}

In summary, we analyzed and computed the statistics of the gap ratio $r$, an essential tool in diagnosing many-body quantum chaos, when
the existence of symmetries results in a block structure of the matrix under consideration. The analytical results we obtain, based on an extension of a seminal work of Rosenzweig and Porter~\cite{Rosenzweig_1960}, are virtually indistinguishable from numerical simulations on large random matrices. While a closed form can only be obtained in limited cases, our formulation, based on Eqs.~\eqref{pxy},\eqref{Hxyfinal} and \eqref{pr2sym}), is compact and generic enough to be implemented easily for all cases of interest. Through several examples of applications, we showed the validity and usefulness of our results to identify or probe symmetries in many-body quantum physics. In this final part of this manuscript, we relate our findings to previous works (including a re-interpretation of results available in the literature) and provide leads for possible extensions.

\subsection{Relation to, and re-interpretation of previous results}
\label{sec:literature}

We now relate our findings to others obtained in studies of spectral statistics in various contexts. Some attempts have been made to count the number of symmetries in chaotic systems \cite{santos2020speck,tekur2018symmetry,bhosale2019superposition}. In Appendix A, we provide a comparison between our results and the techniques proposed to detect symmetries in Ref.~\cite{santos2020speck}.
Our results allow to indirectly discover symmetries in a many-body chaotic system, or to bypass them when they are too complex/costly  to implement numerically. There have been several cases of unusual values of $P(r)$ or $\langle r \rangle$ reported in previous literature which our work directly elucidates. For instance, it applies to the spectral statistics of the Hamiltonian of the fractional quantum Hall effect when orbital inversion is not resolved in the numerics~\cite{Fremling_2018}. Our analysis also explains the results obtained on the 2d square lattice quantum Ising model~\cite{mondaini_eigenstate_2016} in momentum sectors ${\bf k}=(0,0)$ and ${\bf k}=(\pi,\pi)$ where not all symmetries were resolved. The value $P(r=0)\simeq 1.4$ strongly suggests an unresolved $\mathbb{Z}_2$ symmetry there. Our analysis also accounts for the results in the one-dimensional $t-t'-V$ clean fermionic model of  \cite{cheng_many_2016} when the inversion symmetry-breaking field is small, for spectral statistics of the Bose-Hubbard chain~\cite{pausch2020chaos} when the reflection around the center of the chain is not resolved, as well as of quasiperiodic tilings~\cite{roemer} when phase and parity symmetries are not considered. Another context where our work is relevant is the bosonic SYK model with two-body interactions~\cite{iyoda_effective_2018} where the gap ratio distribution (see Fig.~6 in \cite{iyoda_effective_2018}) appears to be close to $P_{m=2}^{\mathrm{GUE}}(r)$, suggesting a two-block GUE structure (for instance due to a particle-hole symmetry), instead of an integrability signature as originally suggested in \cite{iyoda_effective_2018}. For some values of the number of Majorana fermions, the bipartite SYK model introduced in Ref.~\cite{bSYK} displays the average gap ratio value $\langle r \rangle_{2}$ that we derive for the GOE ensemble.

In another direction, our analysis could be useful to discover fracton models~\cite{pretko_fracton_2020,sala_ergodicity_2020,khemani_localization_2020,moudgalya2019thermalization,herviou2020manybody} where the Hamiltonian decomposes in several different Krylov independent blocks (and this not necessarily based on an unresolved symmetry), which necessarily implies a non-adherence to the single-block gap ratio statistics~\cite{atas2013distribution}. A related case is the excellent description of level statistics in an effective quantum ice model~\cite{lee_frustration_2021} with the use of $P_{m=4}^{\mathrm{GOE}}(r)$, accounting for the existing four topological sectors.

\subsection{Perspectives}

Our work can be extended in several directions. In a straightforward way, it is possible to extend the analysis to several blocks with different spectral statistics, for instance, the coexistence of GOE and GUE blocks in the same spectrum. This applies to the quantum Hall work of \cite{Fremling_2018} where different momentum sectors have different spectral statistics. Also, it is possible to see the effect of combining integrable and chaotic blocks, in the spirit of the work of \cite{Berry_1984} on mixed phase spaces. This would apply to models with integrable ``Krylov" subspaces co-existing with ergodic blocks~\cite{moudgalya2019thermalization,sala_ergodicity_2020,khemani_localization_2020}, or to the effective model of the MBL transition proposed in \cite{wei2020characterization} with one ergodic block and random independent energies. 

Our approach is general enough that it should apply mutatis mutandis to other RMT ensembles or other joint distributions $p(s,t)$. Here we considered the Wigner ensembles with quadratic potential in Eq.~\eqref{joinedGOE}. More generally, $\beta$-ensembles with different potentials can be treated in the same way. For instance, $\beta$-Laguerre ensembles, with logarithmic potential, are connected with Wishart matrices and are relevant to characterize entanglement spectra (for a review see e.g. \cite{majumdarextreme}). Entanglement spectra can also exhibit block structure inherited from the symmetry of the underlying quantum state from which they are formed. Other natural extensions include replacing Eq.~\eqref{wignersurmize3}, which is our starting point, by the equivalent expression for matrices of larger sizes: indeed \cite{atas2013joint} obtains, from the exact expression of the joint spacing distribution for $4\times 4$ matrices, an expression for $P(r)$ which is more accurate than Eq.~\eqref{prRMT} by an order of magnitude.
Another possible direction is to study the non-Hermitian situation~\cite{sa_complex_2020} with symmetries. 

A natural generalization would be to consider higher-order spacing ratios: as was shown numerically in \cite{tekur2018symmetry,bhosale2019superposition} higher-order ratios of random matrices with $m$-fold symmetry can be related with ratios of random matrices with Dyson exponent $m$, allowing to detect underlying symmetries. An extension to our work could provide analytical grounds to these observations.

Finally, it would be interesting to analyze the case of weak symmetry breakings (with small matrix elements between different blocks), using a perturbative approach to estimate $P(r)$ in the same vein as the computation performed for the level spacing distribution in \cite{Leitner_1993}.

\begin{acknowledgments}
We thank B. Georgeot, L. Herviou for useful discussions as well as R. Vasseur for earlier insightful correspondence on the two-block structure of the periodic RSOS chain.
This work benefited from the support of the project THERMOLOC ANR-16-CE30-0023-02 and the project COCOA ANR-17-CE30-0024-01 of the French National Research Agency (ANR), and by the French Programme Investissements d'Avenir under the program ANR-11-IDEX-0002-02, reference ANR-10-LABX-0037-NEXT and the  EUR grant NanoX  ANR-17-EURE-0009. We acknowledge the use of HPC resources from CALMIP (grants 2018-P0677, 2019-P0677, 2020-P0677) and GENCI (projects 0070500225 and 0090500225). We use the libraries PETSc~\cite{petsc-efficient,petsc-user-ref}, SLEPc~\cite{slepc-toms,slepc-manual} and Strumpack~\cite{ghysels_efficient_2016,ghysels_robust_2017} for the many-body computations presented in this manuscript.
\end{acknowledgments}

\begin{appendix}
\section{Comparison with other techniques proposed to detect symmetries in chaotic systems}

In view of a practical use of different available methods, we provide in this Appendix elements for a comparison between the approach presented in our work with the methods suggested in Ref.~\cite{santos2020speck} for quantum chaotic systems. 

In Ref.~\cite{santos2020speck} (in particular in its Section VI), two indicators of chaos (the correlation hole and the distribution of off-diagonal elements of local observables) are highlighted  to detect chaos without spectrum unfolding and even in presence of symmetries. Such indicators were introduced and considered in earlier works (see e.g.~Refs.~\cite{Beugeling,leblond,Leviandier,Torres_hole,Schiulaz,delacruz} and references therein).

We first discuss the correlation hole technique, which refers to the existence and detection of a dip in the average survival probability $| \langle \Psi(0) | \Psi(t) \rangle |^2 = | \langle \Psi(0) | \exp(- i Ht) |\Psi(0) \rangle |^2 $ after a quench from an initial state  $|\Psi(0) \rangle$. Results of Ref.~\cite{Schiulaz} indicate that the dip appears after the Thouless time and before the Heisenberg time, which both scale exponentially with system size for many-body systems. The correlation hole is a useful method to detect quantum chaos, and it appears to detect chaos even in the presence of symmetries~\cite{delacruz,Leviandier}. However, this goal is different from the one of the current manuscript, which focuses on positively detecting symmetries using the gap ratio method (and this without the assumption of chaos, as we highlighted).  Furthermore, the two approaches do not have the same computational practicality: the correlation hole method requires to compute the time evolution of the system up to very long times, which in practice means computing {\it all} eigenstates and {\it all} eigenvalues of $H$. This limits this approach to small systems accessible to full diagonalization. Iterative methods, e.g.~using Krylov space techniques, allow to compute the survival probability on larger systems but cannot reach the exponential times required to probe the existence of the correlation hole. On the other hand, the gap ratio method advocated in the present work requires only {\it some} eigenvalues in the middle of the spectrum and {\it no} eigenstates. It is thus amenable to subset methods such as the shift-invert technique~\cite{pietracaprina_shift-invert_2018}, which allows to treat much larger systems (e.g.~matrices of sizes up to $10^7$ in Ref.~\cite{pietracaprina_shift-invert_2018}). 

The second technique discussed in Ref.~\cite{santos2020speck} deals with the distribution of off-diagonal elements of local observables, and consists in computing the ratio $R=\overline{|\langle \alpha|\mathcal{O}|\beta\rangle|^2}/\overline{|\langle \alpha|\mathcal{O}|\beta\rangle|}^2$ between second and first moments of the distribution of off-diagonal elements $\mathcal{O}$ between two different eigenstates $| \alpha \rangle, | \beta \rangle$. Section VI of Ref.~\cite{santos2020speck} suggests (albeit not making any definitive claim) that $R$ can detect symmetries by taking the value $m \pi/2$ when $m$ sectors (of the same size) are present (the value $\pi/2$ comes from the Gaussian distribution of off-diagonal matrix elements in chaotic systems~\cite{Beugeling}). Contrary to the correlation hole method, the $R$ method does not require the full set of eigenstates but only some, and is thus in principle amenable to typically the same system sizes as the gap ratio method. It however suffers from an important drawback: to be useful for symmetry detection, this method requires the observable $\mathcal{O}$ to {\it commute with the symmetry generators}, which means that one needs to know the symmetries in advance.  One could imagine performing trial-and-error by testing different observables in the hope that one of them commutes with the symmetry generators. But this is clearly not error-prone, as we now show by performing computations on the same example as in Ref.~\cite{santos2020speck}: depending on the observable $\mathcal{O}$ we choose, we obtain different results, leading to different conclusions on the number of symmetries/sectors. 
 
\begin{figure}[thp]
\includegraphics[width=\columnwidth]{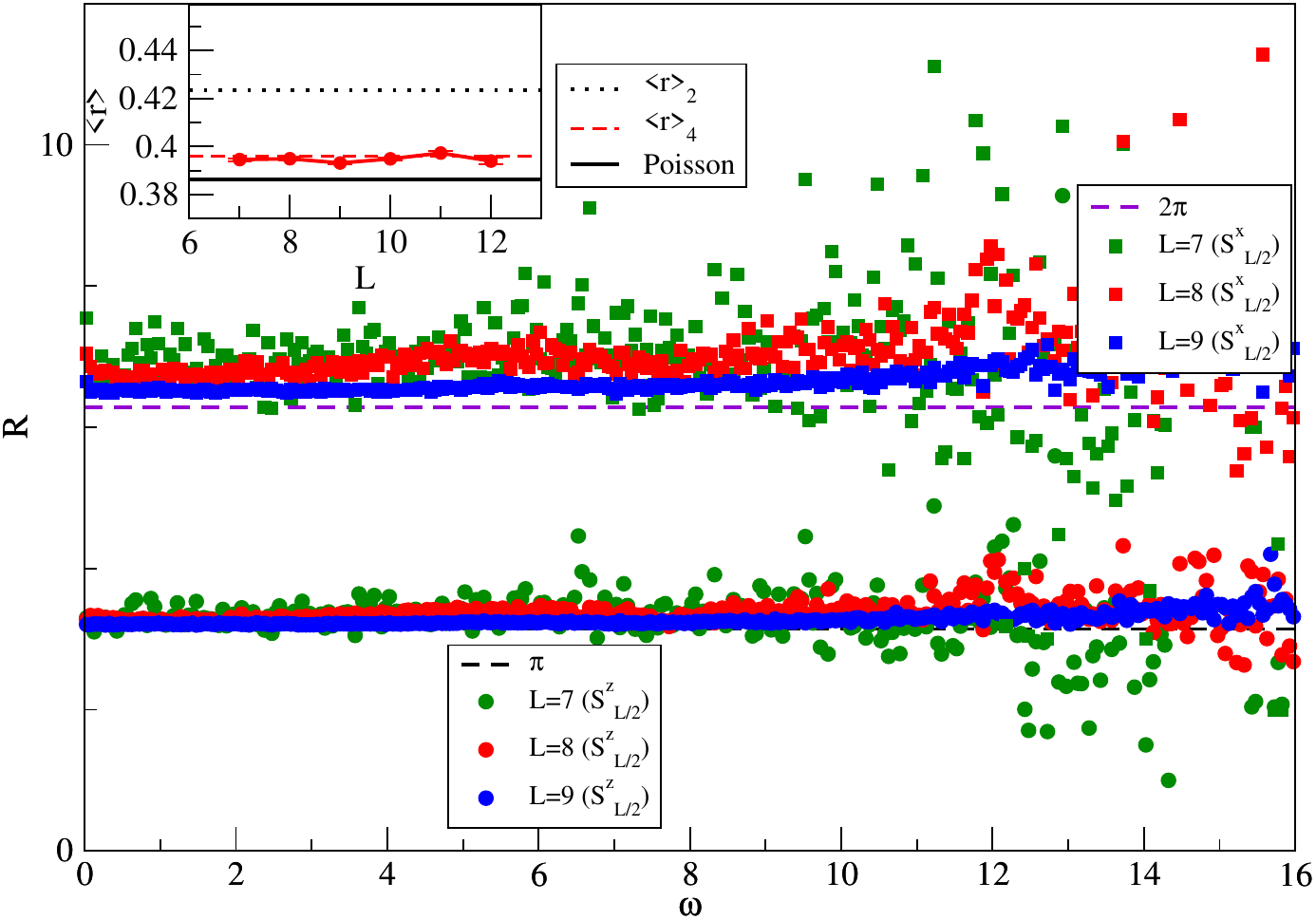}
\caption{$R({\cal O})$ for ${\cal O}=S^z_{L/2}$ (bottom data) and ${\cal O}=S^x_{L/2}$ (top data) for the spin-1 model Eq.~\ref{spin1}, as a function of energy difference $\omega=|E_\alpha - E_\beta|$ between eigenstates $|\alpha\rangle$ and $|\beta\rangle$. Inset: Average gap ratio $\langle r \rangle$ as a function of system size (here $\epsilon=1$).
}
\label{fig:spin1}
\end{figure}

We consider the following one-dimensional $S=1$ spin model
 \begin{eqnarray}
 \label{spin1}
    H_{S1} & = & \sum_{i=1}^{L-1} (S_i^x S_{i+1}^x + S_i^y S_{i+1}^y + S_i^z S_{i+1}^z ) \nonumber \\ 
    &+& \sum_{i=1}^{L-1} (S_i^x S_{i+1}^x)^2+(S_i^y S_{i+1}^y)^2+(S_i^z S_{i+1}^z)^2 \nonumber \\
    & + & \epsilon_1 S_1^x
\end{eqnarray}
(Eq.~3 in Ref.~\cite{santos2020speck}).
The last term is a boundary random magnetic field ($\epsilon_1$ taken uniformly in a box $[-\epsilon,\epsilon]$ with $\epsilon=0.05$) which is added to avoid spatial reflection symmetry, as in Sec.~VI of Ref.~\cite{santos2020speck}. We first consider the same observable ${\cal O}=S_{L/2}^z$ as in Ref.~\cite{santos2020speck}. In that case, as we show in Fig.~\ref{fig:spin1}, we obtain for the ratio $R$ the same flat curves as observed in Fig.~3d in Ref.~\cite{santos2020speck}, with a plateau around $R=\pi$. According to the reasoning of Ref.~\cite{santos2020speck}, this could suggest that there are {\it two} independent sectors. 
If we now compute $R$ for another local observable $\mathcal{O}=S_{L/2}^x$, the data for the corresponding $R$ in Fig.~\ref{fig:spin1} now appear much closer to $2 \pi$, pointing towards $4$ symmetry sectors. The conclusion thus depends on the observable chosen. 
On the contrary, computing the average gap ratio (inset of Fig.~\ref{fig:spin1}, here with $\epsilon=1$ so that the spatial reflection symmetry is strongly suppressed but not changing any symmetry in the model), we obtain $\langle r \rangle \simeq 0.396$, which according to the results in Tab.~\ref{tabcoef} of our manuscript (GOE results since Eq.~\ref{spin1} has real matrix elements) indicates {\it four} sectors. A more thorough analysis of the symmetries in the model \eqref{spin1}~\footnote{F. Alet {\it et al.}, unpublished} confirms that there are indeed four symmetry sectors of identical size.

\end{appendix}

\bibliography{nGOE}

\end{document}